\documentclass[a4paper,12pt]{article}

\usepackage{psfrag}
\usepackage{mathrsfs}

\usepackage{graphicx}
\usepackage[margin=15pt,font=small,labelfont=bf,labelsep=period]{caption}
\usepackage{amsmath}
\usepackage{amssymb}
\usepackage{amsfonts}
\usepackage{mathtools}
\usepackage{bm}
\usepackage{dsfont}
\usepackage{cite}
\usepackage{setspace}
\usepackage{environ,letltxmacro}
\usepackage{tikz}
\usetikzlibrary{arrows,decorations.pathreplacing}
\usepackage{enumitem}
\usepackage{hyperref}
\usepackage{relsize}
\usepackage[a4paper,textwidth=16.5cm,textheight=22.5cm,footskip=15mm]{geometry}

\usepackage{empheq}
\usepackage{environ}
\NewEnviron{boxalign}{\begin{empheq}[box=\fbox]{align} \BODY \end{empheq}}
\usepackage{color}
\usepackage{wasysym}
\usepackage{esint}
\usepackage{subcaption}
\newcommand*\widefbox[1]{\fbox{\hspace{2em}#1\hspace{2em}}} 

\numberwithin{equation}{section}

\LetLtxMacro\oldequation\equation
\LetLtxMacro\endoldequation\endequation
\let\equation\relax
\let\endequation\relax
\NewEnviron{equation}[1][\empty]{%
 \ifx \empty#1
  \oldequation \BODY \endoldequation
 \else
  \ifx b#1
   \oldequation \fbox{$\displaystyle \BODY $} \endoldequation
  \else
   \oldequation \BODY \endoldequation
  \fi
 \fi}

\newcommand{\bg}{\begin{equation}}
\newcommand{\eg}{\end{equation}}

\begin{document}

\begin{titlepage}

\title{
\begin{flushright}
\normalsize{MITP/19-039}
\bigskip
\vspace{1cm}
\end{flushright}
Background Independent Quantum Field Theory and Gravitating Vacuum Fluctuations \\[2mm]
}

\date{}

\author{Carlo Pagani and Martin Reuter\\[3mm]
{\small Institute of Physics (THEP), 
University of Mainz,}\\[-0.2em]
{\small Staudingerweg 7, D--55099 Mainz, Germany}
}

\maketitle
\thispagestyle{empty}

\vspace{2mm}
\begin{abstract}

The scale dependent effective average action for quantum gravity complies
with the fundamental principle of Background Independence. Ultimately
the background metric it formally depends on is selected self-consistently
by means of a tadpole condition, a generalization of Einstein's equation.
Self-consistent backround spacetimes are scale dependent, and therefore
``going on-shell'' at the points along a given renormalization group
(RG) trajectory requires understanding two types of scale dependencies:
the (familiar) direct one carried by the off-shell action functional,
and an equally important indirect one related to the continual re-adjustment
of the background geometry. This paper is devoted to a careful delineation
and analysis of certain general questions concerning the indirect
scale dependence, as well as a detailed explicit investigation of
the case where the self-consistent metrics are determined predominantly
by the RG running of the cosmological constant. Mathematically, the
object of interest is the spectral flow induced by the background
Laplacian which, on-shell, acquires an explicit scale dependence.
Among other things, it encodes the complete information about the
specific set of field modes which, at any given scale, are the degrees
of freedom constituting the respective effective field theory. For
a large class of RG trajectories (Type IIIa) we discover a seemingly
paradoxical behavior that differs substantially from the off-shell
based expectations: A Background Independent theory of (matter coupled)
quantum gravity \emph{looses} rather than gains degrees of freedom
at increasing energies. As an application, we investigate to what
extent it is possible to reformulate the exact theory in terms of
matter and gravity fluctuations on a \emph{rigid flat space}. It turns
out that, in vacuo, this ``rigid picture'' breaks down after a very
short RG time already (less than one decade of scales) because of
a ``scale horizon'' that forms. Furthermore, we critically reanalyze,
and refute the frequent claim that the huge energy densities one obtains
in standard quantum field theory by summing up to zero-point energies
imply a naturalness problem for the observed small value of the cosmological
constant.

\end{abstract}

\end{titlepage}

\newpage

\begin{spacing}{1.1}

\section{Introduction \label{sec:Introduction}}

The cosmological constant presents a conundrum of theoretical physics
that has a very long history already \cite{Weinberg:1988cp,Carroll:2000fy,Straumann:1999ia,Straumann:2002tv}.
The various facets of this problem touch upon both classical and quantum
properties of gravity and matter. For instance, today it is a widely
held opinion that the smallness of the cosmological constant, $\Lambda$,
poses a naturalness problem of unprecedented size. Thereby, according
to one variant of the argument, ``small'' is understood in relation
to the Planck scale, while another one maintains that the energy density
due to $\Lambda$ is unnaturally small in comparison with the vacuum
energy that is believed to result from the quantum field theories
of particle physics. In the present paper we shall mostly be concerned
with the latter version of the ``cosmological constant problem''
and reanalyze it from the point of view of the modern Background Independent
quantum field theory.

\subsection{Summing zero point energies \label{sub:Summing-zero-point-energies}}

The probably best known demonstration of the purported tension between
general relativity and quantum field theory assumes that every mode
of a quantum field on Minkowski space executes zero point oscillations
in the same way the elementary quantum mechanical harmonic oscillator
does, and that this contributes an amount $\frac{1}{2}\hbar\omega$
to the field's ground state energy.

On Minkowski space the modes are characterized by their $3$-momentum
${\bf p}$, and so the total vacuum energy is given by a formal sum
$\sum_{{\bf p}}\frac{1}{2}\hbar\omega\left({\bf p}\right)$. For a
massless free field with $\omega\left({\bf p}\right)=\left|{\bf p}\right|$,
for example, the energy per unit volume is interpreted as the integral
\begin{eqnarray}
\rho_{{\rm vac}} & = & \frac{1}{2}\hbar\int\frac{d^{3}p}{\left(2\pi\right)^{3}}\left|{\bf p}\right|\,,\label{eq:1-1-page6-rho_vac-flat-space}
\end{eqnarray}
which is highly ultraviolet divergent and requires regularization.
One option consists in imposing a sharp momentum cutoff $\left|{\bf p}\right|\leq{\cal P}$
and calculating the integral at finite ${\cal P}$, but other regularization
schemes are possible as well. They all lead to a quartically divergent
energy density
\begin{eqnarray*}
\rho_{{\rm vac}} & = & c{\cal P}^{4}\,,
\end{eqnarray*}
where $c$ is a dimensionless scheme dependent constant of order unity.

Now one fixes ${\cal P}$ at some high scale and argues that $\rho_{{\rm vac}}$
contributes an amount $\Delta\Lambda=8\pi G\rho_{{\rm vac}}=8\pi Gc{\cal P}^{4}$
to the cosmological constant and, as such, ought to be taken into account in Einstein's
equation.

Semiclassical arguments of this kind are presumably due to Pauli \cite{Straumann:1999ia}.
He realized already that even for ${\cal P}$ as low as the familiar
scales of atomic physics the resulting curvature of spacetime becomes
unacceptable should the cosmological constant be of order $\Delta\Lambda$.
Today the cutoff is chosen at the Planck scale often (${\cal P}=m_{{\rm Pl}}$).
Then $\Delta\Lambda$ is about $10^{120}$ times bigger than the observed
cosmological constant, $\Lambda_{{\rm obs}}$. As a result, equating
$\Lambda_{{\rm obs}}$ to a total cosmological constant $\Lambda_{{\rm obs}}=\Lambda_{{\rm bare}}+\Delta\Lambda$
requires fine-tuning $\Lambda_{{\rm bare}}$ at the level of about
$120$ digits.\footnote{Writing the dimensionless ratio $\Lambda/m_{{\rm Pl}}^{2}$ and the
energy density $\rho_{\Lambda}\equiv\Lambda/\left(8\pi G\right)$,
with $m_{{\rm Pl}}\equiv G^{-1/2}$, in the style $\Lambda/m_{{\rm Pl}}^{2}=\Lambda G=\Omega_{\Lambda}h^{2}\times9.15\cdot10^{-122}$
and $\rho_{\Lambda}=\left[\Omega_{\Lambda}^{1/4}h^{1/2}\times3.0\,{\rm meV}\right]^{4}$,
the recently measured $\Omega_{\Lambda}\approx0.68$, $h\approx0.67$
yield the observational values $\Lambda/m_{{\rm Pl}}^{2}\approx2.8\cdot10^{-122}$
and $\rho_{\Lambda}=\left[2.2\,{\rm meV}\right]^{4}$, respectively \cite{Akrami:2018vks}.}

For any plausible choice of the cutoff scale ${\cal P}$ there is
always a flagrant discrepancy between the naturally expected and the
observed values of $\Lambda$. This has nurtured the suspicion that
there could be something fundamentally wrong about the above reasoning.
In the present paper we argue that this is indeed the case.

\subsection{Background Independent QFT}

In this paper we are going to approach the ``cosmological constant
problem'', and in particular the gravitational impact of vacuum fluctuations,
in the light of modern insights from quantum gravity -- even though
the problem is not specifically related to a quantized gravitational
field. In fact, the most profound and, in a sense, even defining difference
between quantum gravity and standard quantum field theory (QFT) is
Background Independence \cite{Ashtekar:2014kba}. This requirement is the key
structural property which we would like to take over from classical
general relativity, ranking higher in fact than, say, questions concerning
the choice of the field variables or the precise form of the dynamics
(field equations, actions).

Whatever approach to quantum gravity one favors (Loop Quantum Gravity,
Causal Dynamical Triangulations, Asymptotic Safety, etc.) the first
and foremost difficulty is always the absence of any pre-exisiting
spacetime geometry that could serve as the ``habitat'' of the dynamical
degrees of freedom. Rather, the geometrical data describing spacetime
(metric, connection, etc.) are themselves subject to quantization.
Hence the highly ``precious'' tool of a spacetime metric, indispensable
in all developments of standard QFT, is available at best on the level
of expectation values only (and in the ``unbroken'' phase not even there). 

Thus, the challenge in setting up a Background Independent
quantum theory of (say, metric) gravity consists in finding a formulation
that does not revert to any rigid, nondynamical metric, that would
play a role analogous to the Minkowski metric in a typical particle
physics context.

In the approach to quantum gravity based upon the gravitational Effective
Average Action \cite{R98}, Background Independence is built into the
formalism by re-interpreting the quantization of a given set of fields
without a distinguished background spacetime as equivalent to simultaneously
quantizing those fields on the totality of all possible backgrounds.
In this sense, a (single) Background Independent quantum field theory
is considered equivalent to an infinite family of background dependent
QFTs. Their members are labeled by the data characterizing the background
spacetime, like the background metric $\bar{g}_{\mu\nu}\left(x\right)$
in the most common case. 

More explicitly, the Effective Average Action (EAA) of metric gravity,
$\Gamma_{k}\left[g_{\mu\nu},\bar{g}_{\mu\nu}\right]$, depends both
on the expectation value $g_{\mu\nu}$ of the metric operator, $\hat{g}_{\mu\nu}$,
and the background metric $\bar{g}_{\mu\nu}$ as an independent second
argument.\footnote{For notational simplicity we suppress the Faddeev-Popov ghosts here.
See \cite{RSBook,Percacci:2017fkn} for a more detailed description.} For every fixed $\bar{g}_{\mu\nu}$, the dynamical metric $\hat{g}_{\mu\nu}$
is quantized on this rigid background by following the familiar lines
of standard QFT. This leads to $\bar{g}_{\mu\nu}$-dependent expectation
values $\langle O\rangle_{\bar{g}}$, in particular the one-point
function $\langle\hat{g}_{\mu\nu}\rangle_{\bar{g}}=g_{\mu\nu}$ which,
after the usual Legendre transformation, becomes an independent field variable,
namely the first argument of $\Gamma_{k}\left[g_{\mu\nu},\bar{g}_{\mu\nu}\right]$.
In general the expectation values $\langle\cdots\rangle_{\bar{g}}$
remember from which member $\bar{g}_{\mu\nu}$ of the family of background
dependent QFTs they come, hence the notation.

\subsection{The standard running and its validity}

In this section we consider a standard effective field theory (EFT)
framework and combine it with the familiar quartic renormalization group (RG) running of the
cosmological constant. Thereafter we point out that the straightforward
use and interpretation of the cosmological constant's RG running is
subtle, and we highlight the conceptual issues that arise.

Let us consider an effective theory defined at the UV scale $k_{{\rm UV}}$.
We denote the EFT action $S_{k_{{\rm UV}}}$ and assume that the most
relevant gravitational part is encoded in the Einstein-Hilbert action,
and that higher curvature terms are negligible at the scale $k_{{\rm UV}}$.
Moreover, we assume that the fluctuations of the metric are strongly
suppressed (we are below the Planck scale). Then, for all purposes,
the action $S_{k_{{\rm UV}}}$ depends on a metric $g_{\mu\nu}$ that
is fixed, and not quantized. At the scale $k_{{\rm UV}}$ the quantized
matter fields, denoted by $\psi$, live on this fixed background $g_{\mu\nu}$.
However, they do not live on an arbitrary background geometry, rather
they live on a specific metric $g_{\mu\nu}$ which is determined by
the EFT field equations, in particular Einstein's equation in our
case. (For concreteness, say, a Friedmann-Lemaitre-Robertson-Walker
metric in cosmological applications.) This defines our UV-EFT.

Now we wish to integrate out some UV modes and lower the EFT cutoff
from $k_{{\rm UV}}$ to $k_{{\rm IR}}$. For the sake of the example,
let us consider a free minimally coupled scalar field. The new EFT
will be obtained from $S_{k_{{\rm UV}}}$ by integrating over the
(covariant) momentum modes between $k_{{\rm UV}}$ and $k_{{\rm IR}}$:
\begin{eqnarray*}
S_{k_{{\rm IR}}} & = & S_{k_{{\rm UV}}}+\frac{1}{2}\mbox{Tr}_{k_{{\rm IR}}}^{k_{{\rm UV}}}\left[\log S_{\Lambda_{k_{{\rm UV}}}}^{\left(2\right)}\right]\,.
\end{eqnarray*}
Here $\mbox{Tr}_{k_{{\rm IR}}}^{k_{{\rm UV}}}$ denotes the partial
trace over the modes with momenta between $k_{{\rm UV}}$ and $k_{{\rm IR}}$.
Focussing on the first term of its derivative expansion, one finds
the following RG running of the cosmological constant:
\begin{eqnarray*}
\Lambda_{k_{{\rm IR}}} & \approx & \Lambda_{k_{{\rm UV}}}+c \left(k_{{\rm IR}}^{4}-k_{{\rm UV}}^{4}\right)\,,
\end{eqnarray*}
where $c$ is a numerical constant.
The UV cosmological constant $\Lambda_{k_{{\rm UV}}}$ can
be fixed by requiring to recover the observed value $\Lambda_{{\rm obs}}$
at very low mass scales, i.e., $\Lambda_{k=0}=\Lambda_{{\rm obs}}$.
This leads to $\Lambda_{k_{{\rm UV}}}=\Lambda_{{\rm obs}}+c k_{{\rm UV}}^{4}$,
and so we retrieve the standard fine-tuning needed to recover the
actual value of the cosmological constant. 

We now point out that the above reasoning has a non-trivial short-coming.
When we lower the EFT cutoff from $k_{{\rm UV}}$ to $k_{{\rm IR}}$
we trace over momentum modes related to a fixed metric. This latter
metric is naturally taken to be the solution of the field equation
of the UV EFT. Once we lowered the cutoff to $k_{{\rm IR}}$ we have
a new EFT action at hand, $S_{k_{{\rm IR}}}$. The EFT action $S_{k_{{\rm IR}}}$
has its own field equations that imply, in principle, a new solution.
By assumption, the most relevant part of $S_{k_{{\rm IR}}}$ is still
encoded in the Einstein-Hilbert action, albeit with a different value
for the cosmological constant now. In the case of the Einstein-Hilbert
action it is easy to relate vacuum solutions of the field equations of $S_{k_{{\rm IR}}}$
and $S_{k_{{\rm UV}}}$, respectively, see section \ref{sub:The-Einstein-Hilbert-case}
for more details. One finds
\begin{eqnarray*}
\left(g_{k_{{\rm IR}}}\right)_{\mu\nu} & = & \frac{\Lambda_{k_{{\rm UV}}}}{\Lambda_{k_{{\rm IR}}}}\left(g_{k_{{\rm UV}}}\right)_{\mu\nu}\,
=\,\frac{\Lambda_{{\rm obs}}+c k_{{\rm UV}}^{4}}{\Lambda_{{\rm obs}}+c k_{{\rm IR}}^{4}}\left(g_{k_{{\rm UV}}}\right)_{\mu\nu}\,.
\end{eqnarray*}
It follows that the natural metric to perform computations at the
scale $k_{{\rm IR}}$ is $g_{k_{{\rm IR}}}$ rather than $g_{k_{{\rm UV}}}$.
This fact alone makes it clear that there must be limitations to
the straightforward application of the standard procedure based on
a fixed background metric.

There is, however, an even more striking consequence of the interpretation
of the RG flow. The RG flow is generated by introducing a cutoff into
the spectrum of the kinetic operator of the scalar field, i.e., the
Laplacian. This Laplacian, too, is built via a fixed background metric,
$\Box_{g}=g^{\mu\nu}D_{\mu}D_{\nu}$. However, the natural fixed background
metrics for $S_{k_{{\rm IR}}}$ and $S_{k_{{\rm UV}}}$, respectively,
differ and their Laplacians are related in a non-trivial way:
\begin{eqnarray*}
\Box_{k_{{\rm IR}}}\,=\, g_{k_{{\rm IR}}}^{\mu\nu}D_{\mu}D_{\nu} 
& = & 
\frac{\Lambda_{k_{{\rm IR}}}}{\Lambda_{k_{{\rm UV}}}}g_{k_{{\rm UV}}}^{\mu\nu}D_{\mu}D_{\nu}
\,=\,\frac{\Lambda_{{\rm obs}}+c k_{{\rm IR}}^{4}}{\Lambda_{{\rm obs}}+c k_{{\rm UV}}^{4}} \, \Box_{k_{{\rm UV}}} \,.
\end{eqnarray*}
It appears then that the standard interpretation of the RG on a fixed
background metric is strongly modified if the metric itself is subject
to a non-negligible induced scale dependence. 

In this work we discuss these issues in detail within the framework of
the gravitational EAA, being the prototype of a Background Independent
approach to non-perturbative quantum gravity.
\vspace{1mm}

The remaining sections of this paper are organized as follows. In
Sections \ref{sec:The-Background-Independent-EAA} and \ref{sec:The-Einstein-Hilbert-example}
we review the basics of the gravitational EAA and discuss a number
of special aspects that will play a role later on. In the ensuing
sections we then develop and apply a number of tools for analyzing
physics predictions encoded in the EAA that become visible only after
``going on-shell'', i.e., specializing for field configurations
that are solutions to the effective field equations, generalizations
of Einstein's equation typically. 

In Section \ref{sec:Self-consistent-Background-Geometries}
we focus on on-shell field configurations such that $\langle\hat{g}_{\mu\nu}\rangle_{\bar{g}}=\bar{g}_{\mu\nu}$,
meaning that $\bar{g}_{\mu\nu}=\left(\bar{g}_{k}^{{\rm sc}}\right)_{\mu\nu}$
is a self-consistent, and hence $k$-dependent background metric.
We use them in order to introduce the ``running'' and the ``rigid''
picture, respectively, two distinguished interpretation schemes for
the (on-shell) RG evolution along a given ``generalized RG trajectory'',
i.e., a scale dependent functional $\Gamma_{k}$ together with a
likewise $k$-dependent background metric. 

In Section \ref{sec:The-spectral-flow} we introduce the concept of
a scale-dependent spectrum along a generalized trajectory of this
kind, and in Sections \ref{sec:The-new-scale-q} and \ref{sec:The-relationship-between-q-and-k}
we describe in detail what this ``spectral flow'' tells us about
the pattern according to which field modes get integrated out while
$\Gamma_{k}$ proceeds along the trajectory. Among other applications,
this yields a precise characterization of the space of degrees of
freedom, $\Upsilon_{{\rm IR}}\left(k\right)$, which, in dependence
on the self-consistent background, are available to the effective
field theory at scale $k$. 

In Section \ref{sec:Application-to-the-CC-problem}, this characterization
is employed for a critical, EAA based reassessment of the above reasoning
about the spacetime curvature caused by vacuum fluctuations. We demonstrate
that this argument breaks down when one tries to embed it into a Background
Independent setting, and that its actual range of applicability is
too restricted to cause a naturalness problem.

\section{The Background Independent Effective Average Action \label{sec:The-Background-Independent-EAA}}

In this section we recall the main properties of the gravitational
Effective Average Action \cite{R98} and elaborate on a number of special
aspects that will prove important later on.

\subsection{Spectra and action functionals}

The EAA is defined in terms of a functional integral over the $c$-number
counterpart of the metric operator, again denoted $\hat{g}_{\mu\nu}$.
The integral is rewritten in terms of a fluctuation variable $h_{\mu\nu}$
which parametrizes the deviation of $\hat{g}_{\mu\nu}$ from $\bar{g}_{\mu\nu}$;
in the simplest case of a linear background split, $\hat{h}_{\mu\nu}=\hat{g}_{\mu\nu}-\bar{g}_{\mu\nu}$.
Starting out from a diffeomorphism invariant bare action $S\left[\hat{g},\cdots\right]$
one adds a gauge fixing term and introduces the corresponding Faddeev-Popov
ghosts $C^{\mu}$ and $\bar{C}_{\mu}$, leading to an integral of
the general form \cite{R98}
\begin{eqnarray}
W_{k}\left[J;\bar{g}\right] & = & \log\int{\cal D}\hat{\varphi}\,\exp\left\{ -S_{{\rm tot}}\left[\hat{\varphi},\bar{g}\right]+\int d^{4}x\sqrt{\bar{g}}J_{i}\hat{\varphi}^{i}-\Delta S_{k}\left[\hat{\varphi},\bar{g}\right]\right\} \,.\label{eq:1-page20-Wk-def}
\end{eqnarray}
Here $\hat{\varphi}\equiv\left(\hat{\varphi}^{i}\right)\equiv\left(\hat{h}_{\mu\nu},C^{\mu},\bar{C}_{\mu},\cdots\right)$
denotes the collection of fields integrated over, with the dots indicating
possible matter fields, and $J\equiv\left(J_{i}\right)$ is a set
of source functions coupled to them. The total action $S_{{\rm tot}}$
comprises the bare one, $S\left[\bar{g}_{\mu\nu}+\hat{h}_{\mu\nu},\cdots\right]$,
as well as the gauge fixing and ghost terms. The cutoff action $\Delta S_{k}$
implements an infrared (IR) cutoff at the mass scale $k$ by giving
a mass $\propto k$ to all normal modes of $\hat{\varphi}$ which
have a $\left(\mbox{covariant momentum}\right)^{2}$ smaller than
$k^{2}$.

At this stage the background metric plays a crucial role. Given a
metric $\bar{g}_{\mu\nu}$, we construct the associated (tensor) Laplacian
$\Box_{\bar{g}}\equiv\bar{g}^{\mu\nu}\bar{D}_{\mu}\bar{D}_{\nu}$,
with $\bar{D}_{\mu}$ the covariant derivative pertaining to the Levi-Civita
connection from $\bar{g}_{\mu\nu}$, and study its eigenvalue problem:
\begin{eqnarray}
-\Box_{\bar{g}}\,\chi_{n}\left(x\right) & = & {\cal E}_{n}\,\chi_{n}\left(x\right)\,.\label{eq:2-page21-eigenvalue-eq-for-Laplacian}
\end{eqnarray}
We expand $\hat{\varphi}$ in terms of the eigen-modes $\left\{ \chi_{n}\right\} $,
i.e., $\hat{\varphi}\left(x\right)=\sum_{n}a_{n}\chi_{n}\left(x\right)$,
so that we could think of the path integral as an integration over
all coefficients, $\int{\cal D}\hat{\varphi}\equiv\prod_{n}\int da_{n}$.
Then, up to a normalization constant, $\Delta S_{k}$ is given by
\begin{eqnarray}
\Delta S_{k} & \propto & k^{2}\sum_{n}\int d^{4}x\sqrt{\bar{g}}R^{\left(0\right)}\left(\frac{{\cal E}_{n}}{k^{2}}\right)\chi_{n}\left(x\right)^{2}\,,\label{eq:3-page22-cutoff-action-expanded}
\end{eqnarray}
where $R^{\left(0\right)}\left(z\right)$ is an essentially arbitrary,
monotonically decreasing function which satisfies $R^{\left(0\right)}\left(0\right)=1$,
and $R^{\left(0\right)}\left(\infty\right)=0$, and which smoothly
``crosses over'' near $z=1$. As a result, the mode $\chi_{n}\left(x\right)$
gets equipped with a nonzero mass term $\propto k^{2}\chi_{n}\left(x\right)^{2}$
if its eigenvalue ${\cal E}_{n}$ is smaller than $k^{2}$, otherwise
it is unaffected. This implements the IR cutoff that will cause the
scale dependence of the EAA. In practice it is convenient to rewrite
(\ref{eq:3-page22-cutoff-action-expanded}) as $\Delta S_{k}=\frac{1}{2}\int d^{4}x\sqrt{\bar{g}}\hat{\varphi}\left(x\right){\cal R}_{k}\hat{\varphi}\left(x\right)$,
without resorting to an explicit mode decomposition, with the pseudo-differential
operator
\begin{eqnarray}
{\cal R}_{k}\left[\bar{g}\right] & = & Z_{k}k^{2}R^{\left(0\right)}\left(\frac{-\Box_{\bar{g}}}{k^{2}}\right)\,.\label{eq:4-page22-cutoff-kernel-structure}
\end{eqnarray}
Here $Z_{k}$ is a matrix in the space of fields which takes care
of their possibly different normalizations.

We emphasize that the eigenvalue condition (\ref{eq:2-page21-eigenvalue-eq-for-Laplacian}),
and hence the spectrum $\left\{ {\cal E}_{n}\left[\bar{g}\right]\right\} $
and the set of eigenmodes, $\left\{ \chi_{n}\left[\bar{g}\right]\left(x\right)\right\} $,
carry a parametric dependence on the background metric. This property
will become pivotal in the later discussion.

Finally, we define the gravitational average action $\Gamma_{k}\left[\varphi;\bar{g}\right]$
as the Legendre transform of $W_{k}\left[J;\bar{g}\right]$ with respect
to all $J_{i}$, at fixed $\bar{g}_{\mu\nu}$, with $\Delta S_{k}\left[\varphi;\bar{g}\right]$
subtracted from it. The EAA depends on the variables ``dual'' to
$J$, the expectation values $\varphi\equiv\langle\hat{\varphi}\rangle\equiv\left(h_{\mu\nu},\xi^{\mu},\bar{\xi}_{\mu},\cdots\right)$.
In particular $h_{\mu\nu}\equiv\langle\hat{h}_{\mu\nu}\rangle=\langle\hat{g}_{\mu\nu}\rangle-\bar{g}_{\mu\nu}=g_{\mu\nu}-\bar{g}_{\mu\nu}$
denotes the expectation value of the metric fluctuation.

Starting from the path-integral formula for $W_{k}$ one can prove
a number of general properties satisfied by $\Gamma_{k}$, such as
BRST- and split-symmetry Ward identities, and one can derive an exact
functional RG equation (FRGE),
\begin{eqnarray}
\partial_{t}\Gamma_{k}\left[\varphi;\bar{g}\right] & = & \frac{1}{2}\mbox{STr}\left[\left(\Gamma_{k}^{\left(2\right)}\left[\varphi;\bar{g}\right]+{\cal R}_{k}\left[\bar{g}\right]\right)^{-1}\partial_{t}{\cal R}_{k}\left[\bar{g}\right]\right]\label{eq:3-page24-FRGE}
\end{eqnarray}
from which it may be computed \cite{W93,Ellwanger:1993mw,Morris:1993qb,Reuter:1993kw}. 
Instead of the pair $\left(h_{\mu\nu},\bar{g}_{\mu\nu}\right)$
one may alternatively use $g_{\mu\nu}$ and $\bar{g}_{\mu\nu}$ as
two independent metric variables. For pure gravity, say, one sets
\begin{eqnarray}
\Gamma_{k}\left[g_{\mu\nu},\bar{g}_{\mu\nu},\xi^{\mu},\bar{\xi}_{\mu}\right] & \equiv & \Gamma_{k}\left[h_{\mu\nu},\xi^{\mu},\bar{\xi}_{\mu};\bar{g}_{\mu\nu}\right]\label{eq:4-page25-column-vs-semicolumn-notation}
\end{eqnarray}
For the functional at $\xi=\bar{\xi}=0$ we write $\Gamma_{k}\left[g_{\mu\nu},\bar{g}_{\mu\nu}\right]\equiv\Gamma_{k}\left[h_{\mu\nu};\bar{g}_{\mu\nu}\right]$.

\subsection{Self-consistent backgrounds}

Leaving the ghosts aside, the action $\Gamma_{k}\left[g_{\mu\nu},\bar{g}_{\mu\nu}\right]$
is the generating functional for the 1PI multi-point correlators of
$\hat{g}_{\mu\nu}$.\footnote{More general composite operators $O\left(\hat{g}_{\mu\nu}\right)$
can be included by coupling them to independent sources 
\cite{Pawlowski:2005xe,Igarashi:2009tj,Pagani:2016pad,Pagani:2017tdr,Pagani:2016dof,Becker:2018quq}.} 
It is an ``off-shell'' quantity, without a direct physical interpretation
away from its critical points. In general, a given pair of metrics
$\left(g_{\mu\nu},\bar{g}_{\mu\nu}\right)$ has no intrinsic meaning
for the physical system by itself: It amounts to a forced situation
where the background $\bar{g}_{\mu\nu}$ is prescribed, and the dynamical
field $\hat{g}_{\mu\nu}$ is coupled to an external source which is
chosen so as to enforce the, likewise prescribed, expectation value
$g_{\mu\nu}=\langle\hat{g}_{\mu\nu}\rangle_{\bar{g}}$.

In order to learn about the state (``vacuum'') the system (``Universe'')
selects dynamically, and wants to be in when it is unperturbed by
external sources $\left(J=0\right)$, one can determine the \emph{self-consistent
background metrics}, $\left(\bar{g}_{k}^{{\rm sc}}\right)_{\mu\nu}$ \cite{Becker:2014pea}.
By definition, when the system is placed in a background of this kind,
the metric develops an expectation value precisely equal to the background:
\begin{eqnarray}
\langle\hat{g}_{\mu\nu}\rangle_{\bar{g}=\bar{g}_{k}^{{\rm sc}}}=\left(\bar{g}_{k}^{{\rm sc}}\right)_{\mu\nu} & \Leftrightarrow & \langle\hat{h}_{\mu\nu}\rangle_{\bar{g}=\bar{g}_{k}^{{\rm sc}}}=0\,.\label{eq:5-page31-def-selfconsistent-background}
\end{eqnarray}
This tadpole condition should be read as an equation for $\bar{g}_{k}^{{\rm sc}}$.
Noting that the modified Legendre transformation from $W_{k}$ to
$\Gamma_{k}$ implies the source-field relation (``effective Einstein
equation'')
\begin{eqnarray}
\frac{1}{\sqrt{\bar{g}\left(x\right)}}\frac{\delta\Gamma_{k}\left[\varphi;\bar{g}\right]}{\delta\varphi^{i}\left(x\right)}+{\cal R}_{k}\left[\bar{g}\right]_{j}^{i}\varphi^{j}\left(x\right) & = & J^{i}\left(x\right)\,,\label{eq:6-page32-effective-field-equations}
\end{eqnarray}
the condition (\ref{eq:5-page31-def-selfconsistent-background}) is
seen to be equivalent to the following \emph{tadpole equation}:
\begin{eqnarray}
\frac{1}{\sqrt{\bar{g}\left(x\right)}}\frac{\delta\Gamma_{k}\left[\varphi;\bar{g}\right]}{\delta h_{\mu\nu}\left(x\right)}\Bigr|_{h=0,\bar{g}=\bar{g}_{k}^{{\rm sc}}} & = & 0\,.\label{eq:7-page32-tadpole-equation-in-selfconst-background}
\end{eqnarray}
In this simplified form it applies to the sector of vanishing ghosts.
In the general case, possibly also including matter, the equation
(\ref{eq:7-page32-tadpole-equation-in-selfconst-background}) gets
coupled to analogous ghost and matter equations \cite{RSBook}.

From equation (\ref{eq:7-page32-tadpole-equation-in-selfconst-background})
it is obvious that the self-consistent backgrounds $\left(\bar{g}_{k}^{{\rm sc}}\right)_{\mu\nu}$
inherit a scale depedence from $\Gamma_{k}$. This fact will become
crucial later on.

\subsection{Bipartite spectra: above and below the cutoff-mode}

We saw that the scale dependent action $\Gamma_{k}\left[g,\bar{g},\cdots\right]$
is intimately related to a family of spectral problems labeled by
$\bar{g}$:
\begin{eqnarray}
-\Box_{\bar{g}}\,\chi_{n}\left[\bar{g}\right]\left(x\right) & = & {\cal E}_{n}\left[\bar{g}\right]\,\chi_{n}\left[\bar{g}\right]\left(x\right)\,.\label{eq:10-page33-eigenvalue-eq-Laplacian-gbar}
\end{eqnarray}
{\bf (1) Eigenbases.} In the limit $k\rightarrow0$, i.e.~when the
IR regulator is removed, the EAA approaches the standard effective
action $\Gamma\left[g,\bar{g},\cdots\right]=\lim_{k\rightarrow0}\Gamma_{k}\left[g_{\mu\nu},\bar{g}_{\mu\nu},\cdots\right]$.
With an eye towards our later discussion we emphasize that the computation
of $\Gamma\left[g,\bar{g},\cdots\right]$, at fixed $\bar{g}_{\mu\nu}$,
really amounts to integrating out \emph{all }the eigenmodes of $\Box_{\bar{g}}$.
Or, stated more explicitly, we integrate over the coefficients $\left\{ a_{n}\right\} $
appearing in the expansion of a generic field $\hat{\varphi}\left(x\right)=\sum_{n}a_{n}\chi_{n}\left[\bar{g}\right]\left(x\right)$
with respect to a complete \emph{basis in field space}, $\left\{ \chi_{n}\left[\bar{g}\right]\right\} $.
Exactly the same remark applies to $\Gamma_{k}$. At non-zero $k$,
the integral (\ref{eq:1-page20-Wk-def}) is over the same domain of
$\hat{\varphi}$'s as for $k=0$; it is only the integrand that changes.\vspace{3mm}\\{\bf (2) The cutoff mode.}
Among the eigenfunctions $\chi_{n}\left[\bar{g}\right]$ there is
one that plays a distinguished role, namely the \emph{cutoff mode},
$\chi_{{\rm COM}}\left[\bar{g}\right]\equiv\chi_{n_{{\rm COM}}}\left[\bar{g}\right]$.
By definition \cite{Reuter:2005bb}, the cutoff mode is the eigenfunction whose
eigenvalue either equals the cutoff scale exactly ${\cal E}_{n_{{\rm COM}}}=k^{2}$,
or, in the case of a discrete spectrum, is the smallest eigenvalue
equal to, or above $k^{2}$, so ${\cal E}_{n_{{\rm COM}}}\gtrsim k^{2}$.

If the eigenvalue with $n=n_{{\rm COM}}\equiv n_{{\rm COM}}\left(k\right)$
is degenerate, there exists actually a set of linearly independent
cutoff modes; it is denoted {\sf COM}$\left(k\right)$.\vspace{3mm}\\{\bf (3) UV vs.~IR-modes.}
When one lowers the cutoff from the ultraviolet $\left(k=\infty\right)$
towards the infrared $\left(k=0\right)$, then for every scale $k$
the mode $\chi_{{\rm COM}}$ is located precisely at the boundary
between \emph{UV-modes}, which have eigenvalues ${\cal E}_{n}\geq{\cal E}_{n_{{\rm COM}}}$
and are integrated out unsuppressed essentially, and the \emph{IR-modes}
with ${\cal E}_{n}\leq{\cal E}_{n_{{\rm COM}}}$; their contribtion
under the functional integral is suppressed by a non-zero regulator
term.

It has to be emphasized that the $k$-dependent division of the eigenbasis
$\Upsilon\left[\bar{g}\right]\equiv\left\{ \chi_{n}\left[\bar{g}\right]\right\} $
into, respectively, an UV-part, which is denoted $\Upsilon_{{\rm UV}}\left[\bar{g}\right]$
and includes the cutoff mode, and an IR-part $\Upsilon_{{\rm IR}}\left[\bar{g}\right]$,
is performed \emph{for each background metric separately}:
\begin{equation}
\boxed{
\Upsilon\left[\bar{g}\right]  =  \Upsilon_{{\rm UV}}\left[\bar{g}\right]\left(k\right)\cup\Upsilon_{{\rm IR}}\left[\bar{g}\right]\left(k\right)\,.
}
\label{eq:11-page35-UVvsIR-mode-partition}
\end{equation}
Assuming that a certain function $\chi\left(x\right)$ happens to
be the eigenfunction of both $-\Box_{\bar{g}_{1}}$ and $-\Box_{\bar{g}_{2}}$,
with eigenvalues ${\cal E}_{1}$ and ${\cal E}_{2}$, respectively,
it is therefore perfectly possible that ${\cal E}_{1}>k^{2}$, but
${\cal E}_{2}<k^{2}$. Thus, at fixed $k$, a given mode function
can very well be classified as of ``UV-type'' when the EAA $\Gamma_{k}\left[g,\bar{g},\cdots\right]$
is evaluated at $\bar{g}=\bar{g}_{1}$, while it is of ``IR-type''
for $\bar{g}=\bar{g}_{2}$.

\begin{figure}
\begin{center}
\includegraphics[scale=0.3]{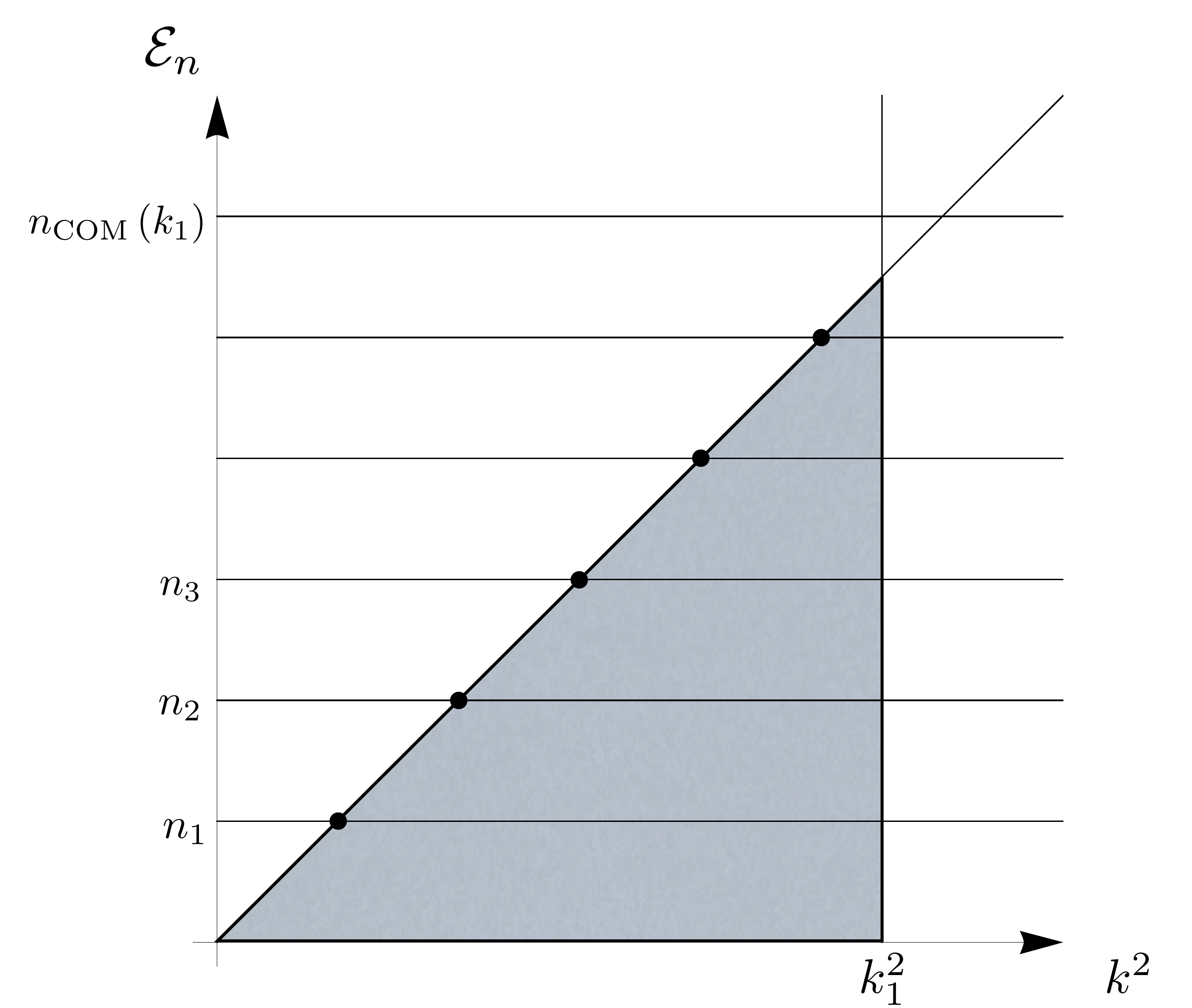}
\end{center}
\caption{Schematic sketch of the trivial ``spectral flow'' of the Laplacian
$-\Box_{\bar{g}}$ for a generic scale-independent background metric
$\bar{g}_{\mu\nu}$. The horizontal lines represent the $k$-independent
eigenvalues, while the diagonal represents the identity map $k^2 \protect\mapsto k^2$.
At the intersection points, a specific mode gets ``integrated out''.
At a given scale $k=k_{1}$, the IR degrees of freedom $\Upsilon_{{\rm IR}}\left[\bar{g}\right]\left(k_{1}\right)$
correspond to precisely those eigenvalues which pass through the shaded
triangle.\label{fig:figure-1}}
\end{figure}

In Figure \ref{fig:figure-1} we illustrate the spectrum of $-\Box_{\bar{g}}$
and the cutoff mode in a style that will prove helpful in the more
complicated situations we shall encounter later on.\vspace{3mm}\\{\bf (4) Importance of $\Upsilon_{\rm{IR}}\left[\bar{g}\right] \left(k\right)$ in effective field theory.}
The decomposition (\ref{eq:11-page35-UVvsIR-mode-partition}) has
a clearcut physical interpretation, which relates to the EAA-based
quantization of the dynamical fields $\hat{\varphi}\equiv\left(\hat{h}_{\mu\nu},\cdots\right)$
on a fixed, i.e., $k$-independent background $\bar{g}_{\mu\nu}$:\begin{enumerate}[label=(\roman*)]
\item For every given scale, $k=k_1$, say, the dynamical impact of all $\hat{\varphi}$-modes in $\Upsilon_{{\rm UV}}\left[\bar{g}\right]\left(k_1\right)$ is encoded in the values of the scale dependent (``running'') coupling constants which parametrize $\Gamma_{k_1} \left[\varphi ; \bar{g}\right]$ at this same scale $k_1$. Or, to use a colloquialism, the UV-modes have been ``integrated out'' already.
\item The vacuum fluctuations of the IR-modes in $\Upsilon_{{\rm IR}}\left[\bar{g}\right]\left(k_1\right)$ are not accounted for by the values of the running couplings at $k_1$. They have not (yet) been integrated out. Using another colloquialism we can say that the functional $\Gamma_{k_1}$ defines an {\it effective field theory} appropriate at the mass scale $k_1$.
\end{enumerate}

The term ``effective field theory'' has many facets \cite{Fredenhagen:2006rv}.
Here, it has only the following simple meaning for us. If one uses
the action functional $\Gamma_{k_{1}}$ rather than the bare action
$S$ in order to compute observables, the only degrees of freedom
that remain to be quantized are those related to the modes in $\Upsilon_{{\rm IR}}\left[\bar{g}\right]\left(k_{1}\right)$.
This is equivalent to saying that \emph{from the perspective of an
effective field theory, the scale $k_{1}$ plays the role of an ultraviolet
cutoff}. All its relevant modes have eigenvalues ${\cal E}_{n}\left[\bar{g}\right]<k^{2}$.

Clearly, one way of ``integrating out'' the modes of $\Upsilon_{{\rm IR}}\left[\bar{g}\right]\left(k_{1}\right)$
is to simply use the FRGE in order to run the RG evolution down to
a lower scale and to let $k_{1}\rightarrow0$ ultimately. However,
in principle the quantization of the IR modes may equally well be
performed by any other technique that allows us to restrict the field
modes to the subset $\Upsilon_{{\rm IR}}\left[\bar{g}\right]\left(k_{1}\right)$.\footnote{Under favourable conditions even perturbation theory might be sufficient.
In fact, when the $\Gamma_{k_{1}}$-theory is really ``effective''
in the usual sense of the word, observables involving a single typical
scale of the order of $k_{1}$ sometimes, but not always, can even
be evaluated without any loop calculations, i.e.~by evaluating $\Gamma_{k_{1}}$
at the classical level. Our present discussion relies in no way on
such special circumstances.}\vspace{3mm}\\{\bf (5) The artificial world ``off-shell''.} To
summarize, we may say that the functional
\[
\left(\varphi;\bar{g}\right)\mapsto\Gamma_{k_{1}}\left[\varphi;\bar{g}\right]
\]
can be thought of as the classical action with a built-in ultraviolet
cutoff at the mass scale $k_{1}$; it governs a reduced, possibly
even finite set of degrees of freedom, $\Upsilon_{{\rm IR}}\left[\bar{g}\right]\left(k_{1}\right)$,
and this set is determined by the spectral problem of the Laplacian
in the respective background geometry, $\Box_{\bar{g}}$. The \emph{numerical
value} of $\Gamma_{k_{1}}\left[\varphi;\bar{g}\right]$ for a given
pair of fields $\left(\varphi;\bar{g}\right)$ is characteristic of
a specific, doubly ``artificial'' situation: First, by unspecified
external means an ad hoc classical metric $\bar{g}_{\mu\nu}$ is installed
on the spacetime manifold, and second, by tuning the external sources
which couple to $\hat{\varphi}=\left(\hat{h}_{\mu\nu},\cdots\right)$,
an expectation value of those fields equal to the prescribed $\varphi$
is enforced, $\langle\hat{\varphi}\rangle=\left(h_{\mu\nu},\cdots\right)=\varphi$.

\section{The Einstein-Hilbert Example \label{sec:The-Einstein-Hilbert-example}}

\begin{figure} 
\begin{subfigure}{.5\textwidth}   
\centering   
\includegraphics[width=.8\linewidth]{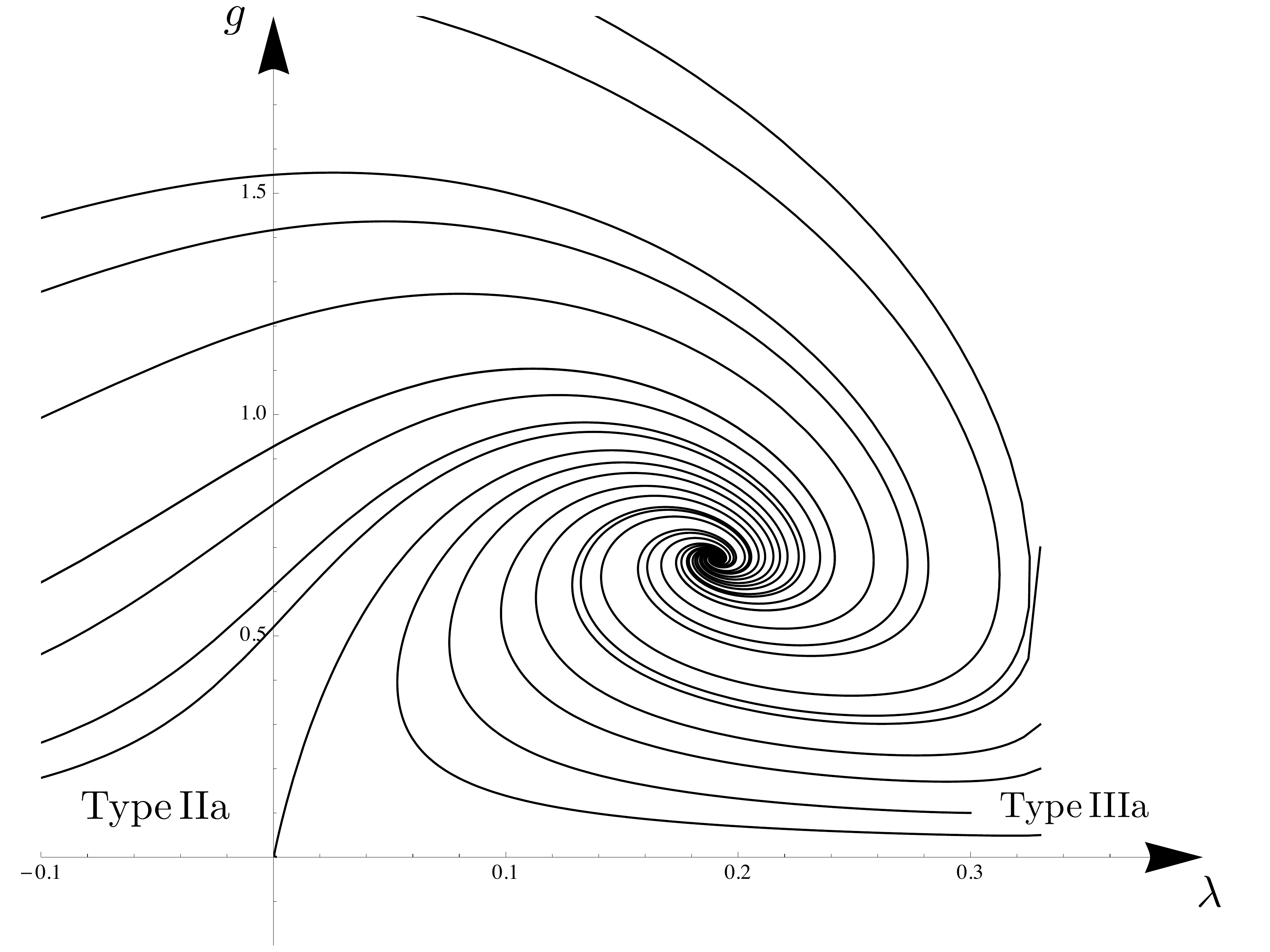}   
\caption{}   \label{fig:sfig1} 
\end{subfigure}%
\begin{subfigure}{.5\textwidth}   
\centering   
\includegraphics[width=1\linewidth]{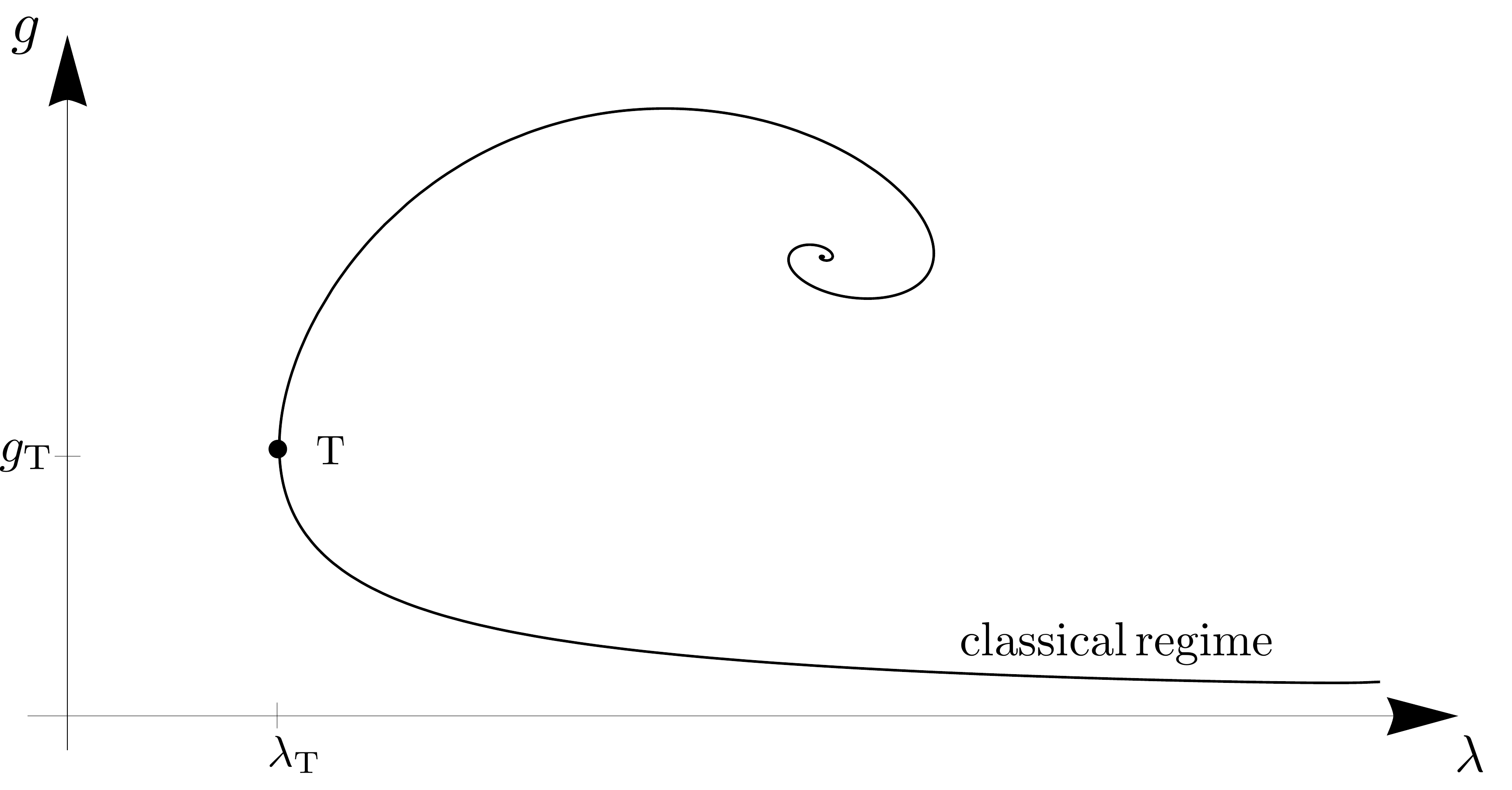}   
\caption{}   \label{fig:sfig2} 
\end{subfigure} 
\caption{
{\bf (a)} Phase portrait of the RG flow on the $\left(g,\lambda\right)$-plane.
The trajectories emanate from the non-Gaussian UV fixed point and flow towards
the IR. 
{\bf (b)} A typical trajectory of Type IIIa. \label{fig:figure2-EH-flow-type-I-II-IIIa}}  
\end{figure}

In the rest of this paper we assume that we are given an (in principle
exact) RG trajectory $k\mapsto\Gamma_{k}\left[g,\bar{g},\xi,\bar{\xi}\right]$.
While our general discussions do not rely on any approximation or
truncation, we shall often invoke the Einstein-Hilbert truncation
as an illustrative example.\vspace{3mm}\\{\bf (1)} The Einstein-Hilbert
truncation of theory space relies on the ansatz
\begin{eqnarray}
\Gamma_{k}\left[g,\bar{g},\xi,\bar{\xi}\right] & = & -\frac{1}{16\pi G_{k}}\int d^{4}x\sqrt{g}\left(R\left(g\right)-2\Lambda_{k}\right)+\cdots\,,\label{eq:20-page47-EH-truncation}
\end{eqnarray}
where the dots stand for the classical gauge fixing and ghost terms.
The RG equations for the dimensionless Newton and cosmological constants,
$g_{k}\equiv G_{k}k^{2}$ and $\lambda_{k}\equiv\Lambda_{k}/k^{2}$,
respectively, are well known \cite{R98}, and their numerical solution \cite{Reuter:2001ag} 
leads to the phase portrait in Figure \ref{fig:figure2-EH-flow-type-I-II-IIIa}.

On the half-plane with $g>0$ we can distinguish the trajectories
of Type Ia, IIa, and IIIa, respectively, which are heading towards
negative, vanishing, and positive values in the infrared, respectively.
In the ultraviolet they emanate from the non-Gaussian fixed point
(NGFP) which renders them asymptotically safe: $\lim_{k\rightarrow\infty}\left(g_{k},\lambda_{k}\right)=\left(g_{*},\lambda_{*}\right)$.\footnote{
We refer the reader to \cite{Gies:2016con,Knorr:2017fus,Gonzalez-Martin:2017gza,Eichhorn:2018akn,deBrito:2018jxt,Falls:2018ylp}
for a partial list of recent results.}

In the sequel we mostly focus on trajectories of the Type IIIa since
conceptually they give rise to the most interesting behavior, and
also because positive values of the cosmological constant are of special
interest phenomenologically.\vspace{3mm}\\{\bf (2)} The trajectories
of Type IIIa are special in that they possess a \emph{turning point
}$\left(g_{{\rm T}},\lambda_{{\rm T}}\right)$ near the origin of
the $g$-$\lambda$- plane. The $\beta$-function of the dimensionless
cosmological constant vanishes there: $\beta_{\lambda}\left(g_{{\rm T}},\lambda_{{\rm T}}\right)=k\frac{d}{dk}\lambda_{k}\Bigr|_{k=k_{{\rm T}}}=0$.
Here $k_{{\rm T}}$ denotes the scale at which the trajectory visits
the point $\left(g_{{\rm T}},\lambda_{{\rm T}}\right)$, see Figure
\ref{fig:figure2-EH-flow-type-I-II-IIIa}.

We are particularly interested in trajectories with $g_{{\rm T}},\lambda_{{\rm T}}\ll1$
which pass very close to the Gaussian fixed point (GFP) located at
$g=\lambda=0$. They spend a long RG time in its vicinity and possess
an extended classical regime.\vspace{3mm}\\{\bf (3)} While it is
straightforward to solve the RG equations numerically, there exists
a convenient analytical approximation for Type IIIa trajectories which
leads to transparent closed-form results often. It is obtained by
linearizing the RG equations\footnote{See eqs.~(4.38) and (4.43) of ref.~\cite{R98}.}
about the GFP. The linear equations are easily solved, and one finds
that the dimensionless couplings evolve according to
\begin{eqnarray}
\lambda_{k} & = & \frac{\Lambda_{0}}{k^{2}}+\nu G_{0}k^{2}+\cdots\label{eq:T1-page49-1-dimless-lambda-linear-flow}\\
g_{k} & = & G_{0}k^{2}+\cdots\,.\label{eq:T2-page49-1-dimless-g-linear-flow}
\end{eqnarray}
The corresponding dimensionful quantities behave as
\begin{eqnarray}
\Lambda_{k} & = & \Lambda_{0}+\nu G_{0}k^{4}+\cdots\label{eq:T3-page49-1-dimful-lambda-linear-flow}\\
G_{k} & = & G_{0}+\cdots\,.\label{eq:T4-page49-1-dimful-g-linear-flow}
\end{eqnarray}
The prefactor $\nu$ in (\ref{eq:T1-page49-1-dimless-lambda-linear-flow})
and (\ref{eq:T3-page49-1-dimful-lambda-linear-flow}) is a constant
of order unity, which depends on the cutoff operator ${\cal R}_{k}$.
For pure gravity it reads
\begin{eqnarray}
\nu & = & \frac{1}{4\pi}\Phi_{2}^{1}\left(0\right)>0\,,\label{eq:T5-page49-2-def-nu-pure-gravity}
\end{eqnarray}
where $\Phi_{2}^{1}$ is one of the standard threshold functions defined
in \cite{R98}.

These solutions amount to a 2-parameter family of RG trajectories
$k\mapsto\left(g_{k},\lambda_{k}\right)$ whose members are labeled
by the constants of integration $\Lambda_{0}$ and $G_{0}$, both
assumed positive. The linear approximation is valid within the classical
and the semiclassical regime of the trajectories. In the former, both
$\Lambda_{k}\approx\Lambda_{0}$ and $G_{k}\approx G_{0}$ are constant
essentially; in the latter, Newton's constant still does not run appreciably,
while $\Lambda_{k}$ is proportional to $k^{4}$. This $k^{4}$ behavior
is reminiscent of the ${\cal P}^{4}$ cutoff dependence mentioned
in Section \ref{sec:Introduction}. As we shall see, their precise
interpretations differ, however.\vspace{3mm}\\{\bf (4)} One easily
checks that the approximate Type IIIa trajectory (\ref{eq:T1-page49-1-dimless-lambda-linear-flow})
and (\ref{eq:T2-page49-1-dimless-g-linear-flow}) indeed possesses
a turning point. Its coordinates are
\begin{eqnarray}
\left(g_{{\rm T}},\lambda_{{\rm T}}\right) & = & \sqrt{G_{0}\Lambda_{0}}\left(\frac{1}{\sqrt{\nu}},2\sqrt{\nu}\right)\,,\label{eq:T10-page49-4-turning-point-lin-flow}
\end{eqnarray}
and it is visited by the trajectory when $k$ assumes the value
\begin{eqnarray}
k_{{\rm T}} & = & \left(\frac{\Lambda_{0}}{\nu G_{0}}\right)^{1/4}\,.\label{eq:T11-page49-4-turning-point-kT}
\end{eqnarray}
Inserting (\ref{eq:T11-page49-4-turning-point-kT}) into (\ref{eq:T3-page49-1-dimful-lambda-linear-flow})
we observe that between $k=0$ and $k=k_{{\rm T}}$ the cosmological
constant increases by precisely a factor of two:
\begin{eqnarray}
\Lambda_{k}\Bigr|_{k=k_{{\rm T}}} & = & 2\Lambda_{0}\,.\label{eq:T-incr-page49-4-Lambda-at-kT}
\end{eqnarray}
\vspace{3mm}\\{\bf (5)} Often it is advantageus to eliminate the
original labels of the trajectories, $\left(G_{0},\Lambda_{0}\right)$,
in favor of the pair $\left(\lambda_{{\rm T}},k_{{\rm T}}\right)$:
\begin{eqnarray}
\Lambda_{0}=\frac{1}{2}\lambda_{{\rm T}}k_{{\rm T}}^{2} & , & G_{0}=\frac{\lambda_{{\rm T}}}{2\nu k_{{\rm T}}^{2}}\,.\label{eq:T12-page49-5-Lambda0-G0-via-lambdaT-kT}
\end{eqnarray}
The relabeling leads to
\begin{eqnarray}
\Lambda_{k} & = & \frac{1}{2}\lambda_{{\rm T}}k_{{\rm T}}^{2}\left[1+\left(\frac{k}{k_{{\rm T}}}\right)^{4}\right]\,=\,\Lambda_{0}\left[1+\left(\frac{k}{k_{{\rm T}}}\right)^{4}\right]\label{eq:T13-page-49-5-Lambdak-linear-via-lamk-kT}\\
G_{k} & = & \frac{\lambda_{{\rm T}}}{2\nu k_{{\rm T}}^{2}}\,=\,\frac{g_{{\rm T}}}{k_{{\rm T}}^{2}}\,.\label{eq:T14-page-49-5-Gk-linea-via-lamk-kT}
\end{eqnarray}
Here $g_{{\rm T}}\equiv\lambda_{{\rm \text{T}}}/\left(2\nu\right)$
is not independent but must be regarded a function of $\lambda_{{\rm T}}$.
The pertinent dimensionelss trajectory writes
\begin{eqnarray}
\lambda_{k} & = & \frac{1}{2}\lambda_{{\rm T}}\left[\left(\frac{k_{{\rm T}}}{k}\right)^{2}+\left(\frac{k}{k_{{\rm T}}}\right)^{2}\right]\label{eq:T15-page49-5-lamk-lin-via-lamT-kT}\\
g_{k} & = & g_{{\rm T}}\frac{k^{2}}{k_{{\rm T}}^{2}}\,.\label{eq:T16-page49-5-gk-lin-via-gT-kT}
\end{eqnarray}

The representation (\ref{eq:T15-page49-5-lamk-lin-via-lamT-kT}) makes
it obvious that the function $k\mapsto\lambda_{k}$ is invariant under
the ``\emph{duality transformation}'' $k\mapsto k_{{\rm T}}^{2}/k$.
As a result, every given value $\lambda>\lambda_{{\rm T}}$ in the
semiclassical regime is realized for two scales, namely $k$ and $k^{\#}=k_{{\rm T}}^{2}/k$,
respectively. (See ref.~\cite{Reuter:2005bb} for a detailed discussion.)\vspace{3mm}\\{\bf (6)}
Once $k$ exceeds a certain critical value, $\hat{k}$, which is of
the order of the ``Planck mass'' $G_{0}^{-1/2}\equiv m_{{\rm Pl}}$,
the linearization about the GFP is no longer a reliable approximation.
The (dimensionless) trajectory enters the scaling regime of the NGFP
then, and ultimately comes to a halt there: $\lim_{k\rightarrow\infty}\left(g_{k},\lambda_{k}\right)=\left(g_{*},\lambda_{*}\right)$.
On the other hand, the dimensionful couplings keep running according
to the power laws
\begin{eqnarray}
\Lambda_{k} & = & \lambda_{*}k^{2}\label{eq:T20-page49-7-Lambdak-at-FP}\\
G_{k} & = & g_{*}k^{-2}\label{eq:T21-page49-7-Gk-at-FP}
\end{eqnarray}
when $k\rightarrow\infty$. \vspace{3mm}\\{\bf (7)} Type IIIa trajectories
displays the following three independent length scales:\begin{enumerate}[label=(\roman*)]
\item The {\it Planck length} as defined by the constant of integration $G_0$,
\begin{equation}
\ell_{\rm{Pl}}\equiv m_{\rm{Pl}}^{-1}=\sqrt{G_0}. 
\label{eq:T30-page49-8-def-Pl-length}
\end{equation}
\item The {\it turning point scale} as determined by $G_0$ and $\Lambda_0$,
\begin{equation}
\ell \equiv k_{\rm{T}}^{-1}=\left( \nu G_0/\Lambda_0 \right)^{1/4}. 
\label{eq:T31-page49-8-def-length-inv-kT}
\end{equation}
\item The {\it Hubble-like length}
\footnote{The name is motivated by the fact that $\sqrt{3/\Lambda_0}$ equals exactly the Hubble length in the case of de Sitter space, and that $\lambda_*=O\left(1\right)$.}
\begin{equation}
L \equiv \left( \lambda_* / \Lambda_0 \right)^{1/2}. 
\label{eq:T32-page49-8-def-Hubble-length}
\end{equation}
\end{enumerate}

In terms of those scales, the running cosmological constant along
a Type IIIa trajectory may be characterised compactly by
\begin{eqnarray}
\Lambda_{k} & = & \Lambda_{0}\times\begin{cases}
1+\left(\ell k\right)^{4} & \mbox{for }0\leq k\apprle\hat{k}\\
\left(Lk\right)^{2} & \mbox{for }k\apprge\hat{k}
\end{cases}\label{eq:T33-page-49-9-Lambdak-via-ell-L}
\end{eqnarray}
with a transition scale $\hat{k}=O\left(m_{{\rm Pl}}\right)$.

Since $\nu,\lambda_{*}=O\left(1\right)$, both hierarchies among these
scales are controlled by the same dimensionless number, namely $G_{0}\Lambda_{0}$:
\begin{eqnarray}
\left(\frac{\ell_{{\rm Pl}}}{\ell}\right)^{4}=\frac{1}{\nu}G_{0}\Lambda_{0} & , & \left(\frac{\ell}{L}\right)^{4}=\frac{\nu}{\lambda_{*}^{2}}G_{0}\Lambda_{0}\,.\label{eq:T34-page49-10-ell-ellPl-L-ratios}
\end{eqnarray}
\vspace{3mm}\\{\bf (8)} With their parameters adjusted accordingly,
the formulae (\ref{eq:T33-page-49-9-Lambdak-via-ell-L}) apply also
to Einstein-Hilbert gravity coupled to a wide variety of matter systems
\cite{Dou:1997fg,Percacci:2002ie,Dona:2013qba,Christiansen:2017cxa,Alkofer:2018fxj},
see \cite{Percacci:2017fkn,Eichhorn:2018yfc} for an overview.
 
When considering matter coupled to gravity in the following
we focus on the subset of matter systems which possess Type IIIa-trajectories
qualitatively similar to those of pure gravity, requiring that
\begin{eqnarray}
\nu,g_{*},\lambda_{*}>0 & \mbox{and} & G_{0},\Lambda_{0}>0\,.\label{eq:T35-page-49-11-positive-inequalities-g-lambda}
\end{eqnarray}
The restrictions (\ref{eq:T35-page-49-11-positive-inequalities-g-lambda})
assumed, the formulae (\ref{eq:T33-page-49-9-Lambdak-via-ell-L})
may be applied to pure and matter coupled gravity alike.

\section{Self-consistent Background Geometries \label{sec:Self-consistent-Background-Geometries}}

In the previous sections we recalled how to define and compute the
Effective Average Action in a Background Independent setting. In this
section we assume instead that we already managed to compute a certain
RG trajectory $k\mapsto\Gamma_{k}$, either via the path integral
or the FRGE. We introduce and analyze a number of scale dependent
objects $\Omega_{k}$ (effective metrics, Laplacians, eigenvalues,
etc.) which ``co-evolve'' with $\Gamma_{k}$, in the sense that,
(i), to compute $\Omega_{k}$ at $k=k_{1}$, the knowledge of $\Gamma_{k_{1}}$
is sufficient, and (ii), the value of $\Omega_{k}$ does not backreact
on the RG evolution $\Gamma_{k}$.\vspace{3mm}\\
{\bf (1)} So from
now on the RG trajectory $k\mapsto\Gamma_{k}$, interpreted as a curve
on theory space, is fixed once and for all. For the essential part
of our discussion it is \emph{not} necessary that this trajectory
is a \emph{complete} one that would in particular be UV-complete and
assign a non-singular function $\Gamma_{k}$ to any $k\in\left[0,\infty\right)$.
The perhaps somewhat surprising phenomena we are going to describe
are relevant even for incomplete trajectories of finite extension,
with $k\in\left[k_{1},k_{2}\right]$, say. Those phenomena are most
pronounced in the semiclassical regime and, logically, they are \emph{unrelated
to all questions of non-perturbative renormalizability}, whether by
Asymptotic Safety any other mechanism.\vspace{3mm}\\{\bf (2)} Given
an RG trajectory $k\mapsto\Gamma_{k}\left[g,\bar{g},\xi,\bar{\xi}\right]$
we can compute the background metrics that are self-consistent at
any point along this trajectory. Focusing on solutions with a vanishing
background value for the ghosts ($\xi=\bar{\xi}=0$), we must solve
the tadpole equation (\ref{eq:7-page32-tadpole-equation-in-selfconst-background})
with $\Gamma_{k}\left[h;\bar{g}\right]\equiv\Gamma_{k}\left[h+\bar{g},\bar{g},0,0\right]$.
In the sequel we assume that this has been done already, and has led
to a certain family of metrics $k\mapsto\left(\bar{g}_{k}^{{\rm sc}}\right)_{\mu\nu}$.\vspace{3mm}\\
So now we must learn how to interpret such explicitly scale dependent
background metrics.
(It is worthy to note that the self-consistent background metrics play also key role
in the computation of the entanglement entropy \cite{Pagani:2018mke}.)

\subsection{From the ``running'' to the ``rigid'' picture \label{sub:From-the-Running-to-rigid-picture}}

To prepare the stage, let us outline how one would extract physical
information from $\Gamma_{k}\left[h;\bar{g}\right]$ or more generally\footnote{In our general considerations we can easily include matter fields
into the discussion replacing $h\rightarrow\left(h,\psi\right)$ everywhere.
Here $\psi$ stands for an arbitrary set of dynamical matter fields.
In fact, in the background field approach to quantum gravity, $h_{\mu\nu}$
is the prototype of a ``matter-like'' field. In the context of the
present paper, $h_{\mu\nu}$ should be regarded logically detached
from $\bar{g}_{\mu\nu}$, together with which it forms the full metric.
Issues of split-symmetry and its breaking \cite{Manrique:2009uh,Manrique:2010mq,Manrique:2010am,Becker:2014qya} will
play no role here}, 
$\Gamma_{k}\left[h,\psi;\bar{g}\right]$, in order to confront its
predictions with with laboratory experiments or cosmological observations.\vspace{3mm}\\
{\bf (1)}
Let us expand the EAA in terms of a $k$-independent set of basis
invariants, $\left\{ I_{\alpha}\left[h,\psi;\bar{g}\right]\right\} $,
and regard it as the generating functional of the running coupling
constants $\bar{u}_{\alpha}\left(k\right)$:\footnote{The overbar of $\bar{u}_{\alpha}\left(k\right)$ indicates that we
are dealing with the \emph{dimensionful} variant of the couplings.}
\begin{eqnarray}
\Gamma_{k}\left[h;\bar{g}\right] & = & \sum_{\alpha}\bar{u}_{\alpha}\left(k\right)I_{\alpha}\left[h,\psi;\bar{g}\right]\,.\label{eq:30-1-page-52-EAA-expansion-I_alpha}
\end{eqnarray}
Now, while it is certainly true that the physics of the (interacting,
non-linear) gravitons $h_{\mu\nu}$ is encoded to some extent in the
$k$-dependent couplings $\bar{u}_{\alpha}\left(k\right)$, that is
only half of the battle. If the dynamically determined background
geometry has a significant $k$-dependence, the invariants $I_{\alpha}$,
evaluated at the physical point $\bar{g}=\bar{g}_{k}^{{\rm sc}}$,
are a second and equally important source of scale dependence:
\begin{eqnarray}
\Gamma_{k}\left[h,\psi;\bar{g}_{k}^{{\rm sc}}\right] & = & \sum_{\alpha}\bar{u}_{\alpha}\left(k\right)I_{\alpha}\left[h,\psi;\bar{g}_{k}^{{\rm sc}}\right]\,.\label{eq:30-2-page-53-EAA-via-I_alpha-sc-background}
\end{eqnarray}

Let us assume that along the RG trajectory there exists an extended
range of $k$-values where the metrics $\bar{g}_{k}^{{\rm sc}}$ are
approximately flat and the running of $\Gamma_{k}$ is negligible,
giving rise to what we call a ``classical regime''. For (notational)
simplicity we assume that this is the case at very low scales near
$k=0$, but the following discussion applies equally well to any other
position of the classical interval on the trajectory.\vspace{3mm}\\
{\bf (2)}
Let us furthermore assume that the large classical Universe at low
scales is inhabited by physicists who are able to perform observations
and experiments both at those low classical scales, and at higher
scales where they perceive a non-trivial RG running already. They
might refer to the former observations as of ``astrophysical'' or
``cosmological'' type, while the latter experiments concern the
``particle physics'' of the $h_{\mu\nu}$- and $\psi$-quanta.

How would these physicists exploit the action (\ref{eq:30-2-page-53-EAA-via-I_alpha-sc-background})
when they try to match it against their observations? First of all
they must take a decision about which metric they prefer to use when
it comes to expressing the values of dimensionful quantities. Natural
options include $\bar{g}_{k}^{{\rm sc}}$ at the running scale $k$,
the metric $\bar{g}_{0}^{{\rm sc}}$ related to the endpoint of the
trajectory, or $\bar{g}_{\kappa}^{{\rm sc}}$ pertaining to any other,
but once and for all fixed scale $k=\kappa$.

Since their macroscopic classical world is well described by $\bar{g}_{0}^{{\rm sc}}$,
the physicists might consider it a sensible starting point to use
$\bar{g}_{0}^{{\rm sc}}$ also when they take first (experimental
and theoretical) steps towards higher scales, with an appreciable
RG running of the metric now. Doing so, it is natural to re-expand
$I_{\alpha}\left[h,\psi;\bar{g}_{k}^{{\rm sc}}\right]$ in terms of
the set $\left\{ I_{\alpha}\left[h,\psi;\bar{g}_{0}^{{\rm sc}}\right]\right\} $
with the $\bar{g}$-argument of all invariants fixed to $\bar{g}_{0}^{{\rm sc}}$.
This set contains the invariants needed to write down a \emph{background-dependent
}theory of gravitons and $\psi$-particles propagating on a rigid
classical spacetime.

Furthermore, one may try to discard the gravitons and use the sub-subset of invariants
$\left\{ I_{\alpha}\left[h=0,\psi;\bar{g}_{0}^{{\rm sc}}\right]\right\} $
in order to formulate a ``standard model of particle physics''.
The metric of spacetime is not an issue then, it is a seemingly universal,
external ingredient, typically the Minkowski metric or $\left(\bar{g}_{0}^{{\rm sc}}\right)_{\mu\nu}=\delta_{\mu\nu}$
in the Euclidean formulation.

The re-expansion of the basis functionals (monomials) is of the form
\begin{eqnarray}
I_{\alpha}\left[h,\psi;\bar{g}_{k}^{{\rm sc}}\right] & = & \sum_{\beta}M_{\alpha\beta}\left(k\right)I_{\beta}\left[h,\psi;\bar{g}_{0}^{{\rm sc}}\right]\label{eq:30-5-page-57-I_alpha-g_k-via-g_0}
\end{eqnarray}
with scale dependent coefficients $M_{\alpha\beta}$. Inserting (\ref{eq:30-5-page-57-I_alpha-g_k-via-g_0})
into (\ref{eq:30-2-page-53-EAA-via-I_alpha-sc-background}), we obtain
a representation of the effective $h$-$\psi$-theory in terms of
$k$-independent basis monomials:
\begin{eqnarray}
\Gamma_{k}\left[h,\psi;\bar{g}_{k}^{{\rm sc}}\right] & = & \sum_{\beta}\left\{ \sum_{\alpha}M_{\alpha\beta}\left(k\right)\bar{u}_{\alpha}\left(k\right)\right\} I_{\beta}\left[h,\psi;\bar{g}_{0}^{{\rm sc}}\right]\,.\label{eq:30-6-page-57-Gamma_k-via-Mab-I_b}
\end{eqnarray}
The equations (\ref{eq:30-2-page-53-EAA-via-I_alpha-sc-background})
and (\ref{eq:30-6-page-57-Gamma_k-via-Mab-I_b}) are two ways of writing
down \emph{the same} functional. Therefore physicists analyzing their
measurements in terms of the rigid metric $\bar{g}_{0}^{{\rm sc}}$
rather than the scale-dependent one, $\bar{g}_{k}^{{\rm sc}}$, actually
do not directly measure the couplings, $\bar{u}_{\alpha}\left(k\right)$,
the natural ones for doing FRGE computations, but the linear
combinations $\sum_{\alpha}M_{\alpha\beta}\left(k\right)\bar{u}_{\alpha}\left(k\right)$.\vspace{3mm}\\
{\bf (3)}
On top of this fairly simple re-organization of the EAA, there is
a second, much more subtle transformation which physicists using no
other metric but $\bar{g}_{0}^{{\rm sc}}$ would want to apply to
$\Gamma_{k}\left[h,\psi;\bar{g}_{k}^{{\rm sc}}\right]$. For them
it appears quite unnatural to parametrize the RG trajectory by the
variable $k$, which is chosen such that $-k^{2}$ is the cutoff in
the spectrum of $\Box_{\bar{g}}$. They will prefer using a new parameter,
$q$, which is likewise connected to a cutoff, but now in the spectrum
of $\Box_{\bar{g}_{0}^{{\rm sc}}}$, the only Laplacian available
to the ``$\bar{g}_{0}^{{\rm sc}}$-only'' physicists.

This raises the nontrivial question of how $q$ is related to the
familiar parameter $k$ which we routinely employ in our FRGE calculations.
What makes this problem particularly intricate is that by evaluating
$\Gamma_{k}$ at $\bar{g}=\bar{g}_{k}^{{\rm sc}}$ the operator $\Box_{\bar{g}}$
itself acquires an explicit $k$-dependence: $\Box_{\bar{g}}\rightarrow\Box_{\bar{g}_{k}^{{\rm sc}}}$.

Before we can address this problem and complete the resulting ``rigid picture'' of the
RG evolution a number of preparatory steps is
needed.

\subsection{The Einstein-Hilbert case \label{sub:The-Einstein-Hilbert-case}}

Within the Einstein-Hilbert truncation the tadpole condition happens
to have the structure of the classical Einstein equation:
\begin{eqnarray}
R_{\mu\nu}\left(\bar{g}_{k}^{{\rm sc}}\right)-\frac{1}{2}R\left(\bar{g}_{k}^{{\rm sc}}\right)\bar{g}_{k,\mu\nu}^{{\rm sc}} & = & 
-\Lambda_{k}\, \bar{g}_{k,\mu\nu}^{{\rm sc}}
+8\pi G_{k}\, T_{k,\mu\nu}\left[\psi_{k}^{{\rm sc}},\bar{g}_{k}^{{\rm sc}}\right]\,.\label{eq:30-page60-Einstein-tadpole-eq}
\end{eqnarray}
The energy-momentum tensor
\begin{eqnarray}
T_{k}^{\mu\nu}\left[\psi,\bar{g}\right]\left(x\right) & \equiv & -\frac{2}{\sqrt{\bar{g}}}\frac{\delta}{\delta h_{\mu\nu} \left(x\right)}\Gamma_{k}^{{\rm M}}\left[h,\psi;\bar{g}\right]\Bigr|_{h=0}\label{eq:30-0-page60-def-EM-tensor}
\end{eqnarray}
makes its appearence only if we generalize the Einstein-Hilbert truncation
ansatz by adding a matter action $\Gamma_{k}^{{\rm M}}\left[h,\psi;\bar{g}\right]$.\footnote{To make the truncation well defined, $\Gamma_{k}^{{\rm M}}$ must
not contain terms $\propto\,\int\sqrt{g}\,,\,\int\sqrt{g}R$.} In that case, (\ref{eq:30-page60-Einstein-tadpole-eq}) gets coupled
to an additional matter field equation, $\delta\Gamma_{k}^{{\rm M}}/\delta\psi=0$.

Throughout this paper, we consider either pure gravity ($\Gamma^{{\rm M}}_k=0\,,\,T_{k}^{\mu\nu}=0$),
or matter-coupled gravity under the simplifying condition that, in
the regime of interest, the $T_{k}^{\mu\nu}$-term on the RHS of (\ref{eq:30-page60-Einstein-tadpole-eq})
is negligible relative to the $\Lambda_{k}$-term:
\begin{eqnarray}
\Lambda_{k} & \gg & 8\pi G_{k}T_{k,\mu}^{\quad\nu}\,.\label{eq:30-1-page60-2-Lambda-bigger-Tmn}
\end{eqnarray}
Here the assumption is that the role played by matter predominantly
consists in renormalizing the pure gravity couplings $\Lambda_{k}$
and $G_{k}$ rather than introducing new ones. (An analogous assumption
is also implicit in Pauli's reasoning.)\vspace{3mm}\\
{\bf (1)}
Thus it will suffice to solve equation (\ref{eq:30-page60-Einstein-tadpole-eq}),
for all $k$ of interest, with $T_{k,\mu\nu}\equiv0$. Clearly this
is still a difficult task in general, but there is a simple way of
promoting any known classical solution (for a fixed cosmological constant)
to a family of $k$-dependent metrics satisfying (\ref{eq:30-page60-Einstein-tadpole-eq}).
The identity $R_{\;\,\,\nu}^{\mu}\left(cg\right)=c^{-1}R_{\;\,\,\nu}^{\mu}\left(g\right)$,
valid for any $c>0$, implies that if $\mathring{g}_{\mu\nu}$ is
a solution of Einstein's classical equation with a fixed cosmological
constant $\mathring{\Lambda}\neq0$, then
\begin{eqnarray}
\left(\bar{g}_{k}^{{\rm sc}}\right)_{\mu\nu} & = & \frac{\mathring{\Lambda}}{\Lambda_{k}}\mathring{g}_{\mu\nu}\label{eq:31-page61-self-const-metric-via-g-fixed}
\end{eqnarray}
is a solution of its $k$-dependent counterpart, eq.~(\ref{eq:30-page60-Einstein-tadpole-eq}),
for any $k$ with $\Lambda_{k}\neq0$. 

According to (\ref{eq:31-page61-self-const-metric-via-g-fixed}),
only the conformal factor of the metric really ``runs'' in this
class of scale dependent background geometries. Obviously the cosmological
constant $\Lambda_{k}$ determines the absolute scale of all lengths
computed from $\bar{g}\equiv\bar{g}_{k}^{{\rm sc}}$ in the expected
way, but it does so differently for changing values of the RG parameter
$k$.\vspace{3mm}\\
{\bf (2)}
Henceforth we employ the approximate analytic formulas for the Type
IIIa trajectory presented earlier. Furthermore, we identify $\mathring{\Lambda}=\Lambda_{k}\Bigr|_{k=0}$
here, but other choices may be natural as well.\footnote{For example, a hypothetical astronomer who is able to measure the
curvature of spacetime on a finite distance scale $1/\kappa<\infty$
would find it convenient to let $\mathring{\Lambda}=\Lambda_{\kappa}$.} Then
\begin{eqnarray}
\left(\bar{g}_{k}^{{\rm sc}}\right)_{\mu\nu} & = & \frac{\Lambda_{0}}{\Lambda_{k}}\left(\bar{g}_{0}^{{\rm sc}}\right)_{\mu\nu}\,=\,Y_{k}^{-1}\left(\bar{g}_{0}^{{\rm sc}}\right)_{\mu\nu}\label{eq:32-page62-g-sc_k-via-Y-g_0}
\end{eqnarray}
where the ratio of cosmological constants,
\begin{eqnarray}
Y_{k} & \equiv & \frac{\Lambda_{k}}{\Lambda_{0}}\,,\label{eq:33-page62-def-Y}
\end{eqnarray}
is given by
\begin{eqnarray}
Y_{k}=1+\ell^{4}k^{4} & , & 0\leq k\apprle\hat{k}\label{eq:34-page62-Y_k-semicl}
\end{eqnarray}
and
\begin{eqnarray}
Y_{k}=L^{2}k^{2} & , & \hat{k}\apprle k<\infty\,,\label{eq:35-page62-FP-regime}
\end{eqnarray}
for the semiclassical and the fixed point regime, respectively.\vspace{3mm}\\
{\bf (3)}
Background metrics of the rescaling type (\ref{eq:31-page61-self-const-metric-via-g-fixed})
entail various simplifications:\begin{enumerate}[label=(\roman*)]
\item We can always organize the basis $\left\lbrace I_\alpha \left[h;\bar{g} \right] \right\rbrace $ in such a way that all $I_\alpha$'s are homogeneous in $\bar{g}_{\mu\nu}$, having a degree of homogeneity $\omega_\alpha$, say. Thus, with (\ref{eq:32-page62-g-sc_k-via-Y-g_0}),
\begin{equation}
I_\alpha \left[h;\bar{g}^{\rm{sc}}_k \right] = Y_k^{-\omega_\alpha} I_\alpha \left[h;\bar{g}^{\rm{sc}}_0 \right]
\label{eq:36-page63-I-of-g_k-via-g_0}
\end{equation}
Hence the re-expansion in eq.~(\ref{eq:30-6-page-57-Gamma_k-via-Mab-I_b}) involves only one term, and $M_{\alpha\beta}=Y_k^{-\omega_\alpha} \delta_{\alpha\beta}$ in this case.
\item The 4D tensor Laplacians $ \Box_{\bar{g}}\equiv \bar{g}^{\mu\nu}\bar{D}_\mu \bar{D}_\nu$ associated with $\bar{g}=\bar{g}^{\rm{sc}}_k$ and $\bar{g}=\bar{g}^{\rm{sc}}_0$, respectively, are related in a simple way,
\begin{equation}
\Box_{\bar{g}^{\rm{sc}}_k} = Y_k \Box_{\bar{g}^{\rm{sc}}_0}
\label{eq:37-page63-Box-g-selfcons-via-Box-g0}
\end{equation}
since
\begin{equation}
\left(\bar{g}^{\rm{sc}}_k \right)^{\mu\nu} = Y_k \left(\bar{g}^{\rm{sc}}_0 \right)^{\mu\nu}
\label{eq:38-page63-g-selfcons-via-g0}
\end{equation}
and the Christoffel symbols, implicit in the Levi-Civita covariant derivative $\bar{D}_\mu$, agree for the two metrics.
\end{enumerate}

\subsection{Maximum symmetry: $S^{4}$-type spaces \label{sub:S4-type-space}}

It will often be instructive to illustrate our general considerations
by means of the technically simplest (Euclidean) background spacetime,
the $4$-sphere $S^{4}\left(\bar{r}\right)$. Its radius $\bar{r}$
is the only free parameter in the metric then, the corresponding line
element being
\begin{eqnarray}
\bar{g}_{\mu\nu}dx^{\mu}dx^{\nu} & = & \bar{r}^{2}ds_{1}^{2}\,,\label{eq:40-page64-1-line-element-S4}
\end{eqnarray}
where $ds_{1}^{2}$ denotes the line element on the round unit sphere,
$S^{4}\left(1\right)$. The metric (\ref{eq:40-page64-1-line-element-S4})
implies the Ricci tensor $R_{\mu\nu}=\left(3/\bar{r}^{2}\right)g_{\mu\nu}$
and the curvature scalar $R=12/\bar{r}^{2}$.\vspace{3mm}\\{\bf (1)}
The spectrum of the Laplacian $\Box_{\bar{g}}$ on $S^{4}\left(\bar{r}\right)$,
acting upon fields of any spin, is well known \cite{Pilch:1984xx,Rubin:1984-A,Rubin:1984tc,Higuchi:1986wu}. The
eigenvalues can be labelled by a single ``quantum number'', $n$,
a positive integer, and one has
\begin{eqnarray}
{\cal E}_{n}=\frac{n\left(n+c_{1}\right)+c_{2}}{\bar{r}^{2}} & , & n=n_{0},n_{0}+1,n_{0}+2,\cdots\,.\label{eq:41-page64-2-Laplacian-spectrum-S4}
\end{eqnarray}
Here $c_{1}$, $c_{2}$, and $n_{0}=0,1,2$ are constants which depend
on the spin, as well as on the dimension, in the general case.

For our present purposes it is sufficient to focus on eigenvalues
with $n\gg1$, leading to a universal formula:
\begin{eqnarray}
{\cal E}_{n}=\left(\frac{n}{\bar{r}}\right)^{2} & , & n\gg1\,.\label{eq:42-page64-2-Laplacian-spectrum-large-n}
\end{eqnarray}
This representation of the $\left(-\Box\right)$-eigenvalues is common
to all fields we shall encounter.\vspace{3mm}\\{\bf (2)} Let us
suppose the sphere $S^{4}\left(\bar{r}_{0}\right)$ is a solution
to the Einstein-Hilbert tadpole equation (\ref{eq:30-page60-Einstein-tadpole-eq})
at $k=0$. This requires $\Lambda_{0}>0$ first of all, and a radius
$\bar{r}_{0}=\sqrt{3/\Lambda_{0}}$. From eq.~(\ref{eq:32-page62-g-sc_k-via-Y-g_0})
we then obtain immediately a family of scale-dependent self-consistent
metrics:
\begin{eqnarray}
\left(\bar{g}_{k}^{{\rm sc}}\right)_{\mu\nu}dx^{\mu}dx^{\nu} & = & 
Y_{k}^{-1}\, \bar{r}_{0}^{2}\, d\bar{s}_{1}^{2}
\,\equiv\,  \bar{r}_{k}^{2}\,  d\bar{s}_{1}^{2}\,.\label{eq:45-page64-3-line-element-S4-self-cons}
\end{eqnarray}
Clearly, at higher scales $k>0$ the spacetimes described by (\ref{eq:45-page64-3-line-element-S4-self-cons})
are still spheres, but with a continually changing radius:
\begin{eqnarray}
\bar{r}_{k} & = & Y_{k}^{-1/2}\bar{r}_{0}\,=\,\sqrt{3/\Lambda_{k}}\,.\label{eq:46-page64-3-k-dependent-radious}
\end{eqnarray}
Typically $\Lambda_{k}$ is an increasing function of $k$, causing
the spacetime to \emph{shrink} at high values of the cutoff scale.

\section{The Spectral Flow \label{sec:The-spectral-flow}}

In Section \ref{sec:The-Background-Independent-EAA} we discussed
the bipartite spectra of $\Box_{\bar{g}}$ for all $k$-independent
backgrounds $\bar{g}$ in the domain of the functional $\Gamma_{k}\left[h;\bar{g}\right]$.
Knowledge of those spectra is necessary in order to \emph{compute}
the EAA. In this section we shall consider a number of related, but
inequivalent spectrum-derived objects in order to \emph{analyze} and
\emph{interpret} an already known trajectory $\Gamma_{k}\left[h;\bar{g}\right]$.

\subsection{Spectral flow induced by the RG trajectory}

To compute $\Gamma_{k}\left[h;\bar{g}\right]$ we had to solve the
eigenvalue problem (\ref{eq:10-page33-eigenvalue-eq-Laplacian-gbar})
for all possible background metrics, at least in principle. Assuming
now that we have the explicit EAA in our hands and try to extract
its physics contents, we are led to go ``on-shell'', i.e., to evaluate
$\Gamma_{k}\left[h;\bar{g}\right]$ and its functional derivatives
at $\bar{g}=\bar{g}_{k}^{{\rm sc}}$.

Therefore our next task is to understand the meaning of the eigenvalue
equation (\ref{eq:10-page33-eigenvalue-eq-Laplacian-gbar}) under
these special circumstances, and to determine the, by now dynamically
selected, space of the effective degrees of freedom, $\Upsilon_{{\rm IR}}$. 

The following algorithm makes it precise what it means to ``insert''
$\bar{g}=\bar{g}_{k}^{{\rm sc}}$ into the UV/IR decomposition (\ref{eq:11-page35-UVvsIR-mode-partition}).
In fact, at first it might appear somewhat confusing that there is
a second source of $k$-dependence now, over and above the ${\rm COM}$-condition
${\cal E}_{n_{{\rm COM}}}=k^{2}$. 
\begin{enumerate}
\item At every \emph{fixed} scale $k$ we freeze the $\bar{g}$-argument
in $\Gamma_{k}\left[h;\bar{g}\right]$, and as a consequence also
in the eigenvalue equation (\ref{eq:10-page33-eigenvalue-eq-Laplacian-gbar}).
This simplifies matters since rather than considering all possible
backgrounds we can now restrict our attention to a single point in
the space of background metrics, namely $\bar{g}_{k}^{{\rm sc}}$.
\item And yet, the overall situation is more involved now since this single
point changes continually when we move along the RG trajectory from
one scale to another. It traces our a certain curve in the space of
metrics: $k\mapsto\bar{g}_{k}^{{\rm sc}}$.
\item The curve of metrics generates an associated curve in the space of
Laplace operators, $k\mapsto\Box_{\bar{g}_{k}^{{\rm sc}}}$. Each
one of those Laplacians, $\Box_{\bar{g}_{k}^{{\rm sc}}}$, gives rise
to its own eigenvalue problem:
\begin{eqnarray}
-\Box_{\bar{g}_{k}^{{\rm sc}}}\,\chi_{n}\left(x;k\right) & = & {\cal F}_{n}\left(k\right)\,\chi_{n}\left(x;k\right)\,.\label{eq:50-page67-spectral-prob-with-g-self-cons}
\end{eqnarray}
At least in principle we can solve (\ref{eq:50-page67-spectral-prob-with-g-self-cons}),
for one value of $k$ after another, and thus obtain a ``curve of
spectra'', or a \emph{spectral flow}, $k\mapsto\left\{ {\cal F}_{n}\left(k\right)\right\} $.
At the same time we find the associated eigenbases, $k\mapsto\left\{ \chi_{n}\left(\,\cdot\,;k\right)\right\} $.
\item Next we determine the respective cutoff modes implied by all spectra that occur
along the curve. At a given point on the curve, having the paramter
value $k=k_{1}$, say, we require that,\footnote{In the case of a discrete spectrum we relax this condition as before:
The cutoff mode possesses the smallest eigenvalue ${\cal F}_{n_{{\rm COM}}}\left(k_{1}\right)$
equal to, or above $k_{1}^{2}$.}
\begin{eqnarray}
{\cal F}_{n_{{\rm COM}}}\left(k\right)\Bigr|_{k=k_{1}} & \stackrel{!}{=} & k_{1}^{2}\,,\label{eq:51-page68-cutoff-mode-with-g-self-cons}
\end{eqnarray}
and we solve this condition for $n_{{\rm COM}}\equiv n_{{\rm COM}}\left(k_{1}\right)$.
In this manner we obtain the label which identifies the cutoff mode
pertaining to the spectrum of the background Laplacian (on-shell!)
at that specific point of theory space which is visited by the RG
trajectory when $k=k_{1}$.
\item Finally we distribute the modes of the eigenbasis $\left\{ \chi_{n}\left(\,\cdot\,;k_{1}\right)\right\} $
over two sets, putting those with with eigenvalues
\begin{eqnarray*}
{\cal F}_{n}\left(k_{1}\right)\geq{\cal F}_{n_{{\rm COM}}\left(k_{1}\right)}\left(k_{1}\right) & \mbox{and} & {\cal F}_{n}\left(k_{1}\right)<{\cal F}_{n_{{\rm COM}}\left(k_{1}\right)}\left(k_{1}\right)
\end{eqnarray*}
into the sets $\Upsilon_{{\rm UV}}\left(k_{1}\right)$ and $\Upsilon_{{\rm IR}}\left(k_{1}\right)$,
respectively.
\end{enumerate}
Repeating this algorithm for all $k_{1}$, we obtain the cutoff mode
for any point of the trajectory, $k\mapsto n_{{\rm COM}}\left(k\right)$,
or more explicitly $k\mapsto\chi_{n_{{\rm COM}}}\left(\,\cdot\,;k\right)$.
Likewise we get the corresponding ``curve'' of UV- and IR- subspaces,
$k\mapsto\Upsilon_{{\rm UV}/{\rm IR}}\left(k\right)$.

In this way we have constructed the decomposition of the eigenbases,
\begin{equation}
\boxed{
\left\{ \chi_{n}\left(\,\cdot\,;k\right)\right\}   =  \Upsilon_{{\rm UV}}\left(k\right)\cup\Upsilon_{{\rm IR}}\left(k\right)
}
\label{eq:52-page69-bipartition-via-spectral-flow}
\end{equation}
which replaces (\ref{eq:11-page35-UVvsIR-mode-partition}) when ``going
on-shell'' and $\bar{g}=\bar{g}_{k}^{{\rm sc}}$ brings in its own
$k$-dependence.
\begin{figure}
\begin{center}
\includegraphics[scale=0.3]{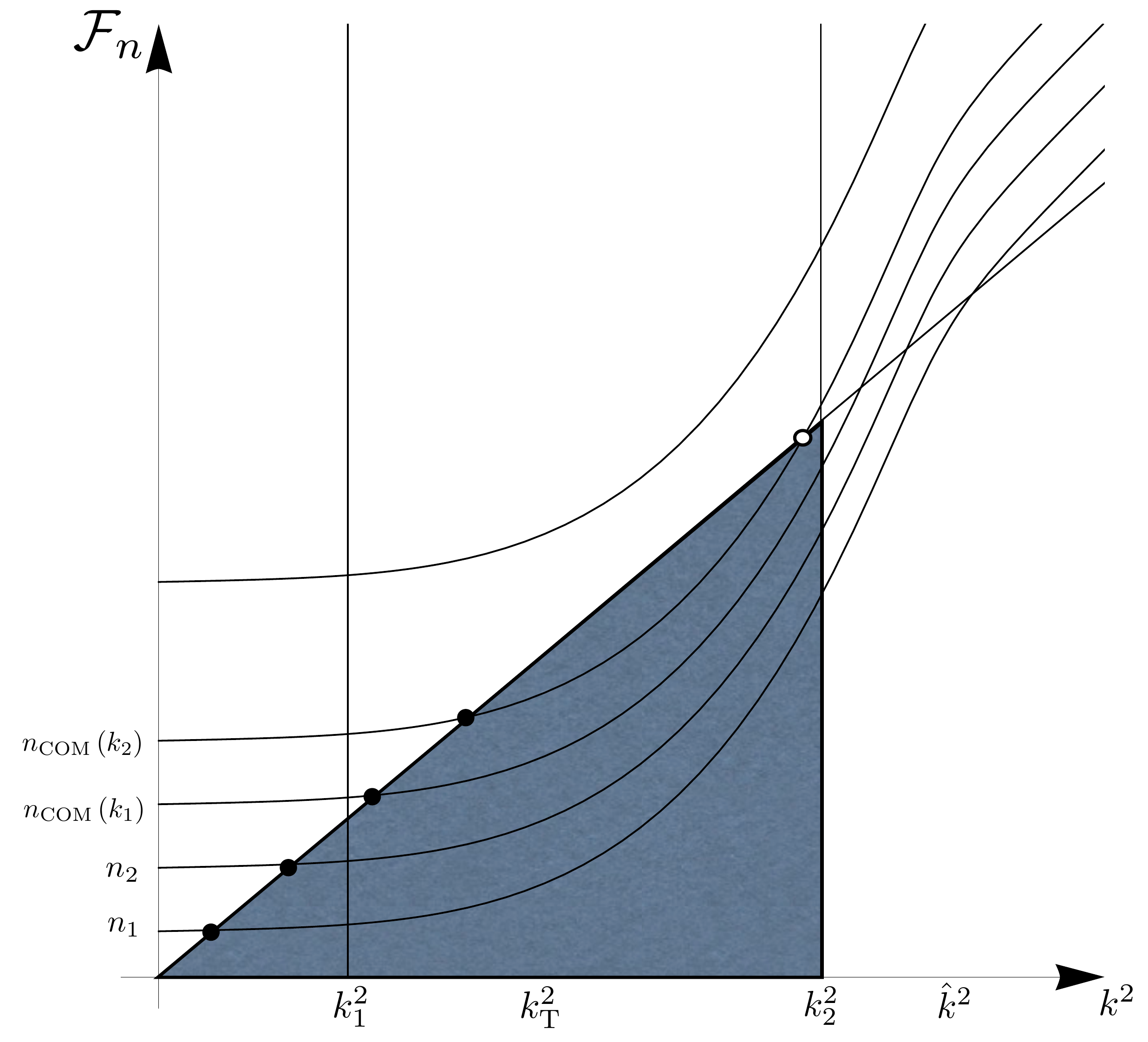}
\end{center}
\caption{Schematic sketch of a non-trivial spectral flow. 
The interpretation is explained in Subsection \ref{subsec:interpretation-spectral-flow} of the text.  \label{fig:figure-3-spectral-flow}}
\end{figure}

Let us now be more explicit and specialize for solutions to the tadpole
equation of the rescaling type (\ref{eq:32-page62-g-sc_k-via-Y-g_0}).

\subsection{Spectral flow for rescaling-type running metrics}

If the $k$-dependence of $\bar{g}_{k}^{{\rm sc}}$ resides in a position-independent
conformal factor only, $\left(\bar{g}_{k}^{{\rm sc}}\right)_{\mu\nu}=Y_{k}^{-1}\left(\bar{g}_{0}^{{\rm sc}}\right)_{\mu\nu}$,
the eigenvalue equation (\ref{eq:50-page67-spectral-prob-with-g-self-cons})
of $-\Box_{\bar{g}_{k}^{{\rm sc}}}$ is solved easily for all $k$
provided its solution is known at a fixed $k$, say $k=0$.

Let us imagine we managed to solve (\ref{eq:50-page67-spectral-prob-with-g-self-cons})
in the special case of $k=0$, and we know all eigenvalues ${\cal F}_{n}\left(k\right)\Bigr|_{k=0}\equiv{\cal F}_{n}^{0}$
and eigenfunctions $\chi_{n}\left(x,k\right)\Bigr|_{k=0}\equiv\chi_{n}^{0}\left(x\right)$:
\begin{eqnarray}
-\Box_{\bar{g}_{0}^{{\rm sc}}}\,\chi_{n}^{0}\left(x\right) & = & {\cal F}_{n}^{0}\,\chi_{n}^{0}\left(x\right)\,.\label{eq:60-page71-Delta-g_0-eigen-equation}
\end{eqnarray}
Multiplying eq.~(\ref{eq:60-page71-Delta-g_0-eigen-equation}) by
$Y_{k}$, and exploiting that $\Box_{\bar{g}_{k}^{{\rm sc}}}=Y_{k}\Box_{\bar{g}_{0}^{{\rm sc}}}$
by (\ref{eq:37-page63-Box-g-selfcons-via-Box-g0}), we obtain
\begin{eqnarray}
-\Box_{\bar{g}_{k}^{{\rm sc}}}\,\chi_{n}^{0}\left(x\right) & = & \left(Y_{k}{\cal F}_{n}^{0}\right)\chi_{n}^{0}\left(x\right)\,.\label{eq:61-page72-Delta_gk-eigen-eq}
\end{eqnarray}
Comparing this relation with eq.~(\ref{eq:50-page67-spectral-prob-with-g-self-cons})
we conclude that, for rescaling-type metrics, the mode functions $\chi_{n}\left(x;k\right)$
are actually independent of $k$, while their eigenvalues possess
a simple scale dependence given by $Y_{k}$:
\begin{subequations}
\begin{empheq}[box=\widefbox]{align}
\chi_{n}\left(x;k\right) & =  \chi_{n}^{0}\left(x\right)\label{eq:62-page72-eigenfunc-chi_k}\\
{\cal F}_{n}\left(k\right) & =  Y_{k}{\cal F}_{n}^{0}\,.\label{eq:63-page72-eigenval-k}
\end{empheq}
\end{subequations}

The multiplicative form of (\ref{eq:63-page72-eigenval-k}) excludes
the possibility of a level crossing, i.e., a re-ordering of the eigenvalues
by the flow. The innocently looking running of eigenvalues in (\ref{eq:63-page72-eigenval-k})
has nevertheless profound implications for the physics of Background
Independent theories, as we discuss next.

\subsection{Interpretation of the spectral flow} \label{subsec:interpretation-spectral-flow}

In Figure \ref{fig:figure-3-spectral-flow} we sketch schematically
a generic spectral flow stemming from a typical RG trajectory $k\mapsto\Gamma_{k}$
along which $Y_{k}=\Lambda_{k}/\Lambda_{0}$ is a rapidly increasing
function of $k$. The horizontal axis corresponds to the trajectory's
curve parameter $k$, while the two vertical lines represent two specific
values of this parameter, $k=k_{1}$ and $k=k_{2}$, respectively.
The presentation in Figure \ref{fig:figure-3-spectral-flow} is analogous
to Figure \ref{fig:figure-1}, whereby ${\cal F}_{n}\left(k\right)$
replaces the constant eigenvalues ${\cal E}_{n}$. The $k$-dependence
of the entire spectrum $\left\{ {\cal F}_{n}\left(k\right)\right\} $
is what we refer to as the spectral flow induced by the RG evolution
of the background metric.\vspace{3mm}\\{\bf (1)} Note that the
scale dependence experienced by the eigenvalues when we move along
the trajectory, \emph{per se}, has nothing to do yet with the concept
of cutoff modes.

In order to determine the cutoff mode for a certain scale, say $k=k_{1}$,
we first locate the points in Figure \ref{fig:figure-3-spectral-flow}
where the graphs of all ${\cal F}_{n}\left(k\right)$ intersect the
vertical line at $k=k_{1}$. Then we check whether the points of intersection
lie above or below the diagonal (${\cal F}=k^{2}$). The UV/IR discrimination
is achieved then by sorting all modes with eigenvalues intersecting, or precisely on the diagonal into the set
$\Upsilon_{{\rm UV}}\left(k_{1}\right)$, and those which intersect
below the diagonal into $\Upsilon_{{\rm IR}}\left(k_{1}\right)$.

The cutoff mode, by definition, is the one with the smallest eigenvalue
among the modes in $\Upsilon_{{\rm UV}}\left(k_{1}\right)$. As indicated
in Figure \ref{fig:figure-3-spectral-flow}, the cutoff mode pertaining
to the specific scale $k=k_{1}$ carries the label $n_{{\rm COM}}\left(k_{1}\right)$.

Clearly, the specific mode with the label $n=n_{{\rm COM}}\left(k_{1}\right)$,
$k_{1}$ fixed, like all the modes, has an eigenvalue ${\cal F}_{n_{{\rm COM}}\left(k_{1}\right)}\left(k\right)$
that depends on the point in the theory space the RG trajectory is
currently visiting, i.e., it depends on $k$ in its role as a curve
parameter. However, for parameter values $k\neq k_{1}$ the mode with
$n=n_{{\rm COM}}\left(k_{1}\right)$ has no special meaning in general.\vspace{3mm}\\{\bf (2)}
As we know, $\Upsilon_{{\rm IR}}\left(k_{1}\right)$ are the modes
not yet integrated out at $k_{1}$, and they consitute the degrees
of freedom governed by the effective action $\Gamma_{k}\Bigr|_{k=k_{1}}$.
In Figure \ref{fig:figure-3-spectral-flow} they are represented by
the eigenvalues passing through the part of the shaded triangle that
lies to the left of the vertical $k=k_{1}$-line. Exactly as in Figure
\ref{fig:figure-1} for the constant ${\cal E}_{n}$, below $k_1$ all those
eigenvalues intersect the diagonal only once; in Figure \ref{fig:figure-3-spectral-flow}
the corresponding intersections points are marked by black circles. 

The physical interpretation of this behavior is deceptively simple:
When we lower $k_{1}$ so that the vertical $k=k_{1}$-line sweeps
over one of the black dots on the diagonal, one more mode is relocated
from $\Upsilon_{{\rm IR}}\left(k_{1}\right)$ into $\Upsilon_{{\rm UV}}\left(k_{1}\right)$.
And, naively, one might think this is exactly as it always must be
since lower cutoffs amount to more modes being ``integrated out''.\vspace{3mm}\\
{\bf (3)}
However, the situation changes profoundly when we move to higher scales,
$k=k_{2}$, in Figure \ref{fig:figure-3-spectral-flow}, say. If the
cosmological constant and $Y_{k}\equiv\Lambda_{k}/\Lambda_{0}$ increase
sufficiently rapidly with $k$, it can happen that below $k_2$ the graph of 
one and the same eigenvalue ${\cal F}_{n}\left(k\right)$ \emph{intersects
the diagonal more than once}. Indeed, Figure \ref{fig:figure-3-spectral-flow}
is inspired by the spectral flow along a Type IIIa trajectory where
this behavior arises as a consequence of the very strong $k^{4}$-running
in the semiclassical regime, see below.

Figure \ref{fig:figure-3-spectral-flow} displays eigenvalues
which both enter and exit the shaded triangle to the left of the $k_{2}$-line; 
the intersection points with the diagonal are marked by black and
open circles, respectively. When we lower $k_{2}$ it may happen that
the vertical $k_{2}$-line sweeps over one of the open circles. Again
this means that a certain mode changed its UV/IR status, but this
time the relocation is from $\Upsilon_{{\rm UV}}\left(k_{2}\right)$
to $\Upsilon_{{\rm IR}}\left(k_{2}\right)$!

At first sight this seems to be a rather strange and perhaps ``unphysical''
phenomenon. After all, we expect that lowering the cutoff leads to
integrating out further modes, and this would move them from $\Upsilon_{{\rm IR}}$
into $\Upsilon_{{\rm UV}}$. Here instead the opposite happens, and
a mode classified ``UV'' all of a sudden becomes ``IR'' by \emph{lowering}
the scale.\vspace{3mm}\\{\bf (4)} This apparent paradox gets resolved,
though, if we recall that the standard equivalence 
$\left(k\mbox{ lowered }\right)\Leftrightarrow \left(\mbox{ mode transfer }\Upsilon_{{\rm IR}}\rightarrow\Upsilon_{{\rm UV}}\right) 
\displaystyle$ holds for the functional $\Gamma_{k}\left[\varphi;\bar{g}\right]$
with $k$-independent (off-shell) field arguments. Importantly, during
the computation of the EAA this equivalence still holds true, also
in the present case. The actual cause of the unexpected spectral behavior
is that when we take the fields on-shell and choose the self-consistent
background they, unavoidably, acquire an extra $k$-dependence which
entails a non-trivial spectral flow then.

A transition $\Upsilon_{{\rm UV}}\rightarrow\Upsilon_{{\rm IR}}$
caused by a lowered $k_{2}$-value is by no means ``unphysical''
therefore. On the contrary, it points to the physically important
fact, that the effective theory given by $\Gamma_{k}$ has \emph{gained}
a new degree of freedom it must deal with; its quantum or statistical
fluctuations are not yet included in the renormalized values of the
couplings comprised by $\Gamma_{k}$.

As Figure \ref{fig:figure-3-spectral-flow} illustrates, at a sufficiently
low scale the new IR-mode crosses the diagonal a second time, thus
leaving the triangle that encompasses the current IR modes.

\subsection{Special cases: Type IIIa and $S^{4}$}

To make the discussion more explicit at this point, we specialize
for Type IIIa trajectories and the maximally symmetric $S^{4}$ solutions
to the Euclidean Einstein equation. \vspace{3mm}\\{\bf (a)} For
a trajectory of Type IIIa all qualitatively essential features are
encapsulated in the simple approximate formulae (\ref{eq:34-page62-Y_k-semicl})
and (\ref{eq:35-page62-FP-regime}) for the semiclassical and the
fixed point regime, respectively. Inserting them into (\ref{eq:63-page72-eigenval-k})
we obtain
\begin{eqnarray}
{\cal F}_{n}\left(k\right) & = & {\cal F}_{n}^{0}\times\begin{cases}
1+\ell^{4}k^{4} & \mbox{for }0\leq k\apprle\hat{k}\\
L^{2}k^{2} & \mbox{for }\hat{k}\apprle k<\infty
\end{cases}\,.\label{eq:80-81page72-10-calF-in-semicl-and-FP-regime}
\end{eqnarray}
The scale dependence of these eigenvalues gives rise to a spectral
flow with exactly the features depicted in Figure \ref{fig:figure-3-spectral-flow}.
The RG effects are strongest in the semiclassical regime. There ${\cal F}_{n}\left(k\right)\propto k^{4}$
increases very rapidly with $k$, and this does indeed lead to eigenvalues
which intersect the diagonal twice.\vspace{3mm}\\{\bf (b)} Opting
for the $S^{4}$-type solutions of the tadpole equation, the generic
label $n$ amounts to a single integer, and we get from equation (\ref{eq:42-page64-2-Laplacian-spectrum-large-n}),
for $n\gg1$, 
\begin{eqnarray}
{\cal F}_{n}\left(k\right)\,=\,\left(\frac{n}{\bar{r}_{k}}\right)^{2} & = & \left(\frac{n}{\bar{r}_{0}}\right)^{2}\times\begin{cases}
1+\ell^{4}k^{4} & \mbox{for }0\leq k\apprle\hat{k}\\
L^{2}k^{2} & \mbox{for }\hat{k}\apprle k<\infty
\end{cases}\,.\label{eq:85-86-page72-11-calF-large-n-semicl-and-FP}
\end{eqnarray}
In particular in the semiclassical regime, the $k$-dependent radius
of the sphere, $\bar{r}_{k}=Y_{k}^{-1/2}\bar{r}_{0}$, decreases rapidly
for increasing $k$, thus causing a corresponding growth of the eigenvalues:
\begin{eqnarray}
\bar{r}_{k} & = & \frac{\bar{r}_{0}}{\sqrt{1+\ell^{4}k^{4}}}\quad\mbox{(semiclassical)}\,.\label{eq:90-page72-11-r_k-semiclass}
\end{eqnarray}
In the fixed point regime, where
\begin{eqnarray}
\bar{r}_{k} & = & \frac{\bar{r}_{0}}{L}\frac{1}{k}\quad\mbox{(fixed point)}\label{eq:91-page72-11-r_k-FP}
\end{eqnarray}
the self-consistent value of the radius decreases more slowly with
$k$. Note also that in the analogous Lorentzian setting $\bar{r}_{k}$
corresponds to the inverse Hubble parameter, i.e., the Hubble length.

\section{The new RG parameter $q$ \label{sec:The-new-scale-q}}

Now we are prepared to return to the physicists living at $k=0$,
who would like to reformulate the entire effective theory (\ref{eq:30-6-page-57-Gamma_k-via-Mab-I_b})
in terms of $\bar{g}_{0}^{{\rm sc}}$. As we mentioned already, this
involves reparametrizing the (fixed!) RG trajectory that is under
scrutiny, $k\mapsto\Gamma_{k}$, in terms of a new scale parameter
$q=q\left(k\right)$.

\subsection{Introducing $q$ as a cutoff scale}

Ideally, in analogy with $k^{2}$ which is a cutoff in the spectrum
of $-\Box_{\bar{g}}\Bigr|_{\bar{g}=\bar{g}_{k}^{{\rm sc}}}$, the
new parameter $q^{2}$ should be an eigenvalue cutoff for the operator
$-\Box_{\bar{g}}\Bigr|_{\bar{g}=\bar{g}_{0}^{{\rm sc}}}$. The latter
has no scale dependence, and it is the only Laplacian the ``$\bar{g}_{0}^{{\rm sc}}$-only''
physicists want to use.\footnote{Again we emphasize that here we are \emph{analyzing} a given RG trajectory,
rather than \emph{computing} it. We are not proposing any different
off-shell functional $\Gamma_{k}\left[h;\bar{g}\right]$ here. In
particular the $q$- and the $k$-schemes, respectively, are not different
ways of expanding the integration variable under the path integral.
In fact, in the process of computing the functional $\Gamma_{k}\left[h;\bar{g}\right]$,
those two cases are completely indistinguishable, as we neither set
$\bar{g}=\bar{g}_{k}^{{\rm sc}}$ nor $\bar{g}=\bar{g}_{0}^{{\rm sc}}$
at that stage; we rather keep $\bar{g}$ fully arbitrary, but independent
of any scale.}

In principle, the division of the eigenfunctions in UV-modes and IR-modes,
respectively, can be described without recourse to any metric, namely
by characterzing the cutoff mode and the sets $\Upsilon_{{\rm UV}/{\rm IR}}$
directly in terms of the mode labels $n$. Usually $n$ is chosen
to be a dimensionless multi-index (one or several ``quantum numbers'')
that is not linked to any particular metric.

Considering running metrics of the rescaling type, equation (\ref{eq:62-page72-eigenfunc-chi_k})
tells us that the eigenfunctions have no explicit scale dependence
despite the running of the metric. Therefore, if a certain function
$\chi_{n}^{0}\left(x\right)\Bigr|_{n=n_{{\rm COM}}}$ is the cutoff
mode at $k=0$, it is so also at any other point along the curve of
spectra induced by the running background. This is true regardless
of whether we parametrize the curve by the standard parameter $k$
or by the new variable $q$. The difference between the $k$- and
$q$-scheme, respectively, arises only when we convert the label $n=n_{{\rm COM}}$
to the (dimensionful) square of a covariant momentum.

By considering eq.~(\ref{eq:63-page72-eigenval-k}) at $n=n_{{\rm COM}}$
we obtain
\begin{eqnarray}
{\cal F}_{n_{{\rm COM}}}\left(k\right) & = & Y_{k}\,{\cal F}_{n_{{\rm COM}}}^{0}\,.\label{eq:70-page76-calF-nCOM-k}
\end{eqnarray}
In the $k$-scheme, the quantum number $n_{{\rm COM}}$ is converted
to a momentum, $k$, by setting
\begin{eqnarray}
{\cal F}_{n_{{\rm COM}}}\left(k\right) & \stackrel{!}{=} & k^{2}\label{eq:71-page76-calF-nCOM-is-k-square}
\end{eqnarray}
and solving for $n_{{\rm COM}}=n_{{\rm COM}}\left(k\right)$. Now
we convert \emph{the same} quantum number $n_{{\rm COM}}$ to the
momentum $q^{2}$ preferred by the ``$k=0$-physicists''. In complete
analogy with (\ref{eq:71-page76-calF-nCOM-is-k-square}) they set
\begin{eqnarray}
{\cal F}_{n_{{\rm COM}}}^{0} & \stackrel{!}{=} & q^{2}\,.\label{eq:72-page76-calF_0-is-q-square}
\end{eqnarray}
Recall that ${\cal F}_{n_{{\rm COM}}}\left(k\right)$ and ${\cal F}_{n_{{\rm COM}}}^{0}$
denote the, in general different, eigenvalues of $-\Box_{\bar{g}_{k}^{{\rm sc}}}$
and $-\Box_{\bar{g}_{0}^{{\rm sc}}}$, respectively, belonging to
their common eigenfunction $\chi_{n_{{\rm COM}}}^{0}\left(x\right)$.
If we now insert (\ref{eq:71-page76-calF-nCOM-is-k-square}) and (\ref{eq:72-page76-calF_0-is-q-square})
into (\ref{eq:70-page76-calF-nCOM-k}) we obtain
\begin{eqnarray}
k^{2} & = & Y_{k}\,q^{2}\,.\label{eq:73-page77-k-square-via-q-square}
\end{eqnarray}
Solving for $q$ yields the new RG parameter as a function of the
old one:
\begin{equation}
\boxed{
q^{2}\left(k\right)  =  \frac{k^{2}}{Y_{k}}\,.
} \label{eq:74-page77-q-square-via-k-square}
\end{equation}
This simple, yet fully explicit formula will be a crucial tool in
the following. In fact, there are two immediate applications in which
the function $k\mapsto q\left(k\right)$ and its inverse play a prominent
role, as we discuss next.

\subsection{Completing the rigid picture: $k=k\left(q\right)$ \label{sub:Completing-the-rigid-picture}}

Let us assume for a moment that it is possible to invert the relation
(\ref{eq:74-page77-q-square-via-k-square}) and obtain $k$ as a function
of $q$, i.e., $k=k\left(q\right)$. Under this assumption we can
finally complete our task of recasting the EAA from (\ref{eq:30-6-page-57-Gamma_k-via-Mab-I_b})
in a style which eliminates $\bar{g}_{k}^{{\rm sc}}$ everywhere in
favour of $\bar{g}_{0}^{{\rm sc}}$. This is achieved by setting
\begin{eqnarray}
\Gamma_{q}\left[h,\psi\right] & \equiv & \Gamma_{k}\left[h,\psi;\bar{g}_{k}^{{\rm sc}}\right]\Bigr|_{k=k\left(q\right)}\,.\label{eq:75-page78-EAA-via-g_0}
\end{eqnarray}
This variant of the EAA refers the physics of all ``particles'',
the graviton included, to one metric only, namely the one that is
self-consistent at $k=0$.

\subsection{Cutoff modes: relating $n_{{\rm COM}}\left(k\right)$ to $q=q\left(k\right)$
\label{sub:Cutoff-modes:-relating-nCOM-and-q}}

In order to determine the cutoff modes along the spectral flow, i.e.,
the $k$-dependent ``quantum number'' $n_{{\rm COM}}\left(k\right)$,
we need information about, first, the RG trajectory, and, second,
about the structure of spacetime. The former information is encoded
in the relationship $q=q\left(k\right)$ and its inverse, while the
latter is provided by the (single) spectrum $\left\{ {\cal F}_{n}^{0}\right\} $
of the scale independent Laplacian $-\Box_{\bar{g}_{0}^{{\rm sc}}}$.
Given these data, eq.~(\ref{eq:72-page76-calF_0-is-q-square}) yields
the condition
\begin{eqnarray}
{\cal F}_{n_{{\rm COM}}\left(k\right)}^{0} & = & q\left(k\right)^{2}\label{eq:76-page79-calF_0-via-q}
\end{eqnarray}
which is to be solved for $n_{{\rm COM}}=n_{{\rm COM}}\left(k\right)$
then. 

For instance, in the case of the $S^{4}$ spacetime, the relevant
spectrum reads ${\cal F}_{n}^{0}=\left(n/\bar{r}_{0}\right)^{2}$,
and so
\begin{eqnarray}
n_{{\rm COM}}\left(k\right) & = & \bar{r}_{0}q\left(k\right)\,.\label{eq:77-page79-nCOM-via-q-of-k}
\end{eqnarray}
It is assumed here that $n\gg1$, so that $n$ is a quasi-continuous
variable and we can be cavalier as for its integer character. 

Note also that since $\bar{r}_{k}=Y_{k}^{-1/2}\bar{r}_{0}$ and $q\left(k\right)=kY_{k}^{-1/2}$,
the result (\ref{eq:77-page79-nCOM-via-q-of-k}) is equivalent to
\begin{equation}
\boxed{
n_{{\rm COM}}\left(k\right)  =  \bar{r}_{0}q\left(k\right)\,=\,\bar{r}_{k}k\,.
}\label{eq:78-page80-nCOM-via-r_k}
\end{equation}
This formula nicely illustrates how the two natural perspectives on
the RG flow are connected, namely the conventional ``running picture'' which employs
the scale parameter $k$ leading to a running spacetime metric, and
the new ``rigid picture'' based on the scale $q$ together with the rigid metric
$\bar{g}_{0}^{{\rm sc}}$. In the $S^{4}$ example, the correspondence
($\bar{r}_{0}\leftrightarrow\bar{r}_{k}$ and $q\leftrightarrow k$)
allows us to interpret \emph{one and the same }mode, labelled $n_{{\rm COM}}$,
as either $q\bar{r}_{0}$, or as $k\bar{r}_{k}$, respectively, depending
on whether the ``rigid'' or the ``running'' perspective
is adopted.

\section{The Global Relation Between $q$ and $k$ \label{sec:The-relationship-between-q-and-k}}

\subsection{Non-invertibility of $q=q\left(k\right)$}

\begin{figure}
\begin{center}
\includegraphics[scale=0.3]{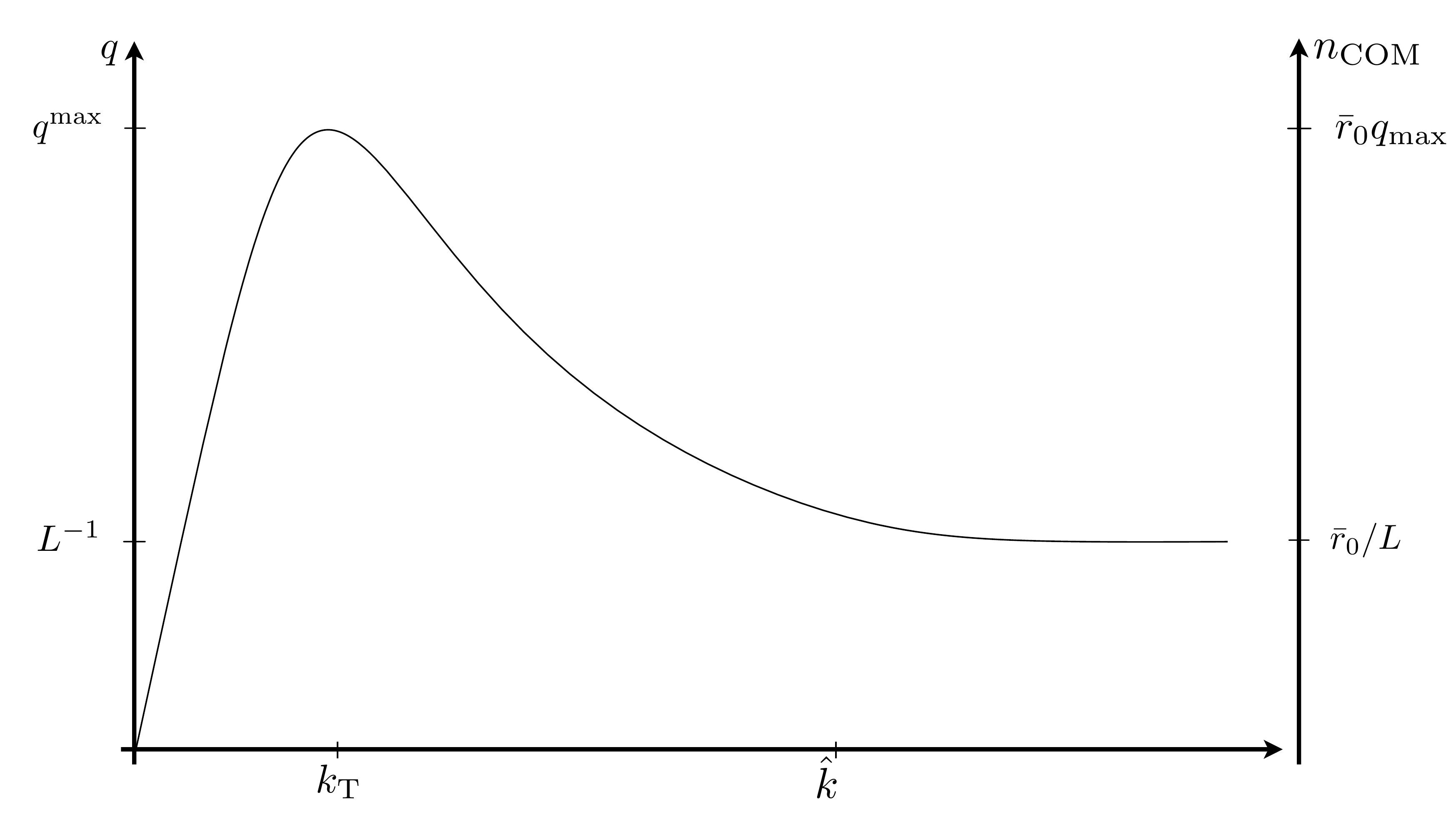}
\end{center}
\caption{The functions $q\left(k\right)$ and $n_{{\rm COM}}\left(k\right)=\bar{r}_{0}q\left(k\right)$
along a Type IIIa trajectory as given by eqs.~(\ref{eq:100-page81-1-q_k-in-semicl})
and (\ref{eq:101-page81-1-q_k-in-FP}).\label{fig:figure-4-q-plot}}
\end{figure}

Along the RG trajectory of the Type IIIa, the ratio of the cosmological
constants, $Y_{k}$, is well approximated by (\ref{eq:34-page62-Y_k-semicl})
and (\ref{eq:35-page62-FP-regime}). This turns eq.~(\ref{eq:74-page77-q-square-via-k-square})
into
\begin{eqnarray}
q\left(k\right) & = & \left(\frac{k^{2}}{1+\ell^{4}k^{4}}\right)^{1/2}\mbox{ if }0\leq k\apprle\hat{k}\label{eq:100-page81-1-q_k-in-semicl}\\
q\left(k\right) & = & L^{-1}\qquad\qquad\;\,\,\,\,\mbox{ if }\hat{k}\apprle k<\infty\,.\label{eq:101-page81-1-q_k-in-FP}
\end{eqnarray}
A plot of the function $k\mapsto q\left(k\right)$ is shown in Figure
\ref{fig:figure-4-q-plot}.

Let us follow a Type IIIa trajectory from the IR towards the UV, see
Figure \ref{fig:figure2-EH-flow-type-I-II-IIIa}b. At all scales far
below the turning point, $k\ll k_{{\rm T}}=\ell^{-1}$, it is in the
classical regime, quantum effects are negligible, and $q\approx k$
approximately. Then, at $k=k_{{\rm T}}$ the trajectory passes the
turning point, the $k^{4}$-running of the cosmological constant sets
in, and once $k\gg k_{{\rm T}}$, eq.~(\ref{eq:100-page81-1-q_k-in-semicl})
yields roughly
\begin{eqnarray}
q\left(k\right) & \approx & \frac{1}{\ell^{2}k}\,\,\,\mbox{ if }k_{{\rm T}}\ll k\ll\hat{k}\,.\label{eq:105-page82-q_k-between-kT-hat_k}
\end{eqnarray}
This behavior is rather striking: An \emph{increasing }value of the
standard RG parameter $k$ implies a \emph{decrease} of the newly
introduced scale $q$. In fact, the function $q\left(k\right)$ has
a maximum precisely at the trajectory's turning point, $k=k_{{\rm T}}=\ell^{-1}$,
where it assumes the value
\begin{eqnarray}
q_{{\rm max}} & = & q\left(k_{{\rm T}}\right)\,=\,\frac{1}{\sqrt{2}}k_{{\rm T}}\,.\label{eq:106-page82-q_max-via-kT}
\end{eqnarray}
For all the other points in the semiclassical regime, eq.~(\ref{eq:100-page81-1-q_k-in-semicl})
always associates two $k$-values to a given $q<q_{{\rm max}}$, see
Figure \ref{fig:figure-4-q-plot}.

Obviously the function $k\mapsto q\left(k\right)$ is not monotonic,
and eq.~(\ref{eq:100-page81-1-q_k-in-semicl}) cannot be inverted
in the entire domain of interst. Therefore, globally speaking, the
map $k\mapsto q\left(k\right)$ is not an acceptable reparametrization
of the RG trajectory $k\mapsto\Gamma_{k}$; it fails to establish
a diffeomorphism on the RG-time axis.

Nevertheless, \emph{locally}, namely for either $k<k_{{\rm T}}$ or
$k>k_{{\rm T}}$, eq.~(\ref{eq:100-page81-1-q_k-in-semicl}) can
be inverted, yielding the following two branches of $k=k\left(q\right)$
for $q\in\left[0,q_{{\rm max}}\right]$:
\begin{eqnarray}
k_{\pm}\left(q\right) & = & \frac{1}{\sqrt{2}\ell^{2}q}\left[1\pm\sqrt{1-\left(2\ell^{2}q^{2}\right)^{2}}\right]^{1/2}\,=\,\sqrt{2}\,\frac{q_{{\rm max}}^{2}}{q}\left[1\pm\sqrt{1-\left(\frac{q}{q_{{\rm max}}}\right)^{4}}\right]^{1/2}\,.\label{eq:107-page83-k_pm-branches-before-kT}
\end{eqnarray}
For $q$ given, the functions $k_{+}\left(q\right)$ and $k_{-}\left(q\right)$
return values smaller and larger than $k_{{\rm T}}=\ell^{-1}$, respectively.
They are joined at $k_{\pm}\left(q_{{\rm max}}\right)=\sqrt{2}q_{{\rm max}}=k_{{\rm T}}$.

If we follow the Type IIIa trajectory beyond $\hat{k}=O\left(m_{{\rm Pl}}\right)$
we enter the asymptotic fixed point regime. The cosmological constant
scales as $\Lambda_{k}=\lambda_{*}k^{2}$ there, hence $Y_{k}\propto k^{2}$,
and as a consequence,  $q^{2}=k^{2}/{\cal F}_{k}$ becomes perfectly
independent of $k$ asymptotically:
\begin{eqnarray}
q\left(k\right) & = & L^{-1}\mbox{ if }k\gg\hat{k}\,.\label{eq:110-page85-q_k-for-k-in-FP-regime}
\end{eqnarray}
Note that since $\lambda_{*}=O\left(1\right)$, the length parameter
$L=\sqrt{\lambda_{*}/\Lambda_{0}}$ is essentialy the radius of the
Universe (Euclidean Hubble length) according to the $\bar{g}_{0}^{{\rm sc}}$-metric.
Hence the asymptotic value $q\left(\infty\right)^{2}=L^{-2}=\Lambda_{0}/\lambda_{*}=3/\left(\lambda_{*}\bar{r}_{0}^{2}\right)$
is an extremely tiny momentum square, even by the standards of the
$k=0$-physicists employing the rigid metric. For them, the Universe
is a sphere of radius $\bar{r}_{0}$, and $q\left(\infty\right)^{2}$
is of the same order of magnitude as the lowest lying eigenvalues
${\cal E}_{n}$ in (\ref{eq:41-page64-2-Laplacian-spectrum-S4}) for
the normal modes on this sphere.

In the present paper, the Asymptotic Safety based ultraviolet completion
of Quantum Einstein Gravity plays no important role. We focus here
on the implications of the ``strange'' relationship between the
two alternative RG scales $k$ and $q$ which we summarize in Figure
\ref{fig:figure-4-q-plot}. Its salient properties are entirely due
to the $k^{4}$-running in the semiclassical regime. In the following
we restrict the discussion to this regime mostly.

\subsection{Scale horizon and the failure of the rigid picture \label{sub:Scale-horizon-and-failure-of-q-picture}}

By the very construction of the FRGE, the IR cutoff scale $k$ provides
a globally valid parametrization of the RG trajectories, $k\mapsto\Gamma_{k}$.
We tried to introduce a new RG-time parameter $q$ such that the size
of all (dimensionful) eigenvalues is expressed relative to the rigid
metric $\bar{g}_{0}^{{\rm sc}}$ rather than the running one, $\bar{g}_{k}^{{\rm sc}}$.
Now we see this ``rigid metric''-persepective on the RG flow is
doomed to fail. 

This perspective appears to be the natural one for the $k=0$-physicists,
who are either ignorant of the RG running in the gravitational sector,
or try to incorporate the quantum gravity effects into the couplings
of an effective action which, however, is still conservative in relying
on fields that live on the same classical spacetime at all RG times.\vspace{3mm}\\
{\bf (1)}
As the function $q=q\left(k\right)$ cannot be inverted globally on
the RG-time axis, it is clear that the reparametrized action $\Gamma_{q}\left[h,\psi\right]$
of eq.~(\ref{eq:75-page78-EAA-via-g_0}) can make sense at best locally.
For example, it does yield a consistent description for small momenta
$q\in\left[0,q_{{\rm max}}\right]$, for which the ``minus'' branch
of (\ref{eq:107-page83-k_pm-branches-before-kT}), $k=k_{-}\left(q\right)$,
establishes a one-to-one relation between $q$ and $k$. In a way,
since $q=0\Leftrightarrow k=0$ here, this amounts to the ``perturbative''
branch of $k\left(q\right)$.

However, starting out at $k=0$ and then following the Type IIIa trajectory
from the IR towards the UV, we come to a point where the $q$-parametrization
breaks down, namely $q=q_{{\rm max}}=k_{{\rm T}}/\sqrt{2}$. This
value is equivalent to $k=k_{{\rm T}}$, i.e., precisely when the
trajectory passes through its turning point, the $q$-description
becomes untenable: Moving from the turning point further towards the
UV,\footnote{Or in unambiguous, global terms: ``Increasing $k$ from $k=k_{{\rm T}}$
to $k=k_{{\rm T}}+\delta k$ with $\delta k>0$, ... .''} while still adhering to the parameter $q$, it would have to decrease
rather than increase, thus conveying the utterly false impression
of an RG evolution which runs in the wrong direction. \vspace{3mm}\\
{\bf (2)}
We conclude that the trajectory's turning point acts as a kind of
horizon for the $k=0$-physicist who try to ignore the RG evolution
of the spacetime geometry as long as possible. The rigid metric perspective
on the Background Independent RG flow comes at the price of a {\it``scale horizon''} 
on the RG time axis (rather than spacetime) beyond which
strong quantum effects render it inconsistent. 

\begin{figure}
\begin{center}
\includegraphics[scale=0.3]{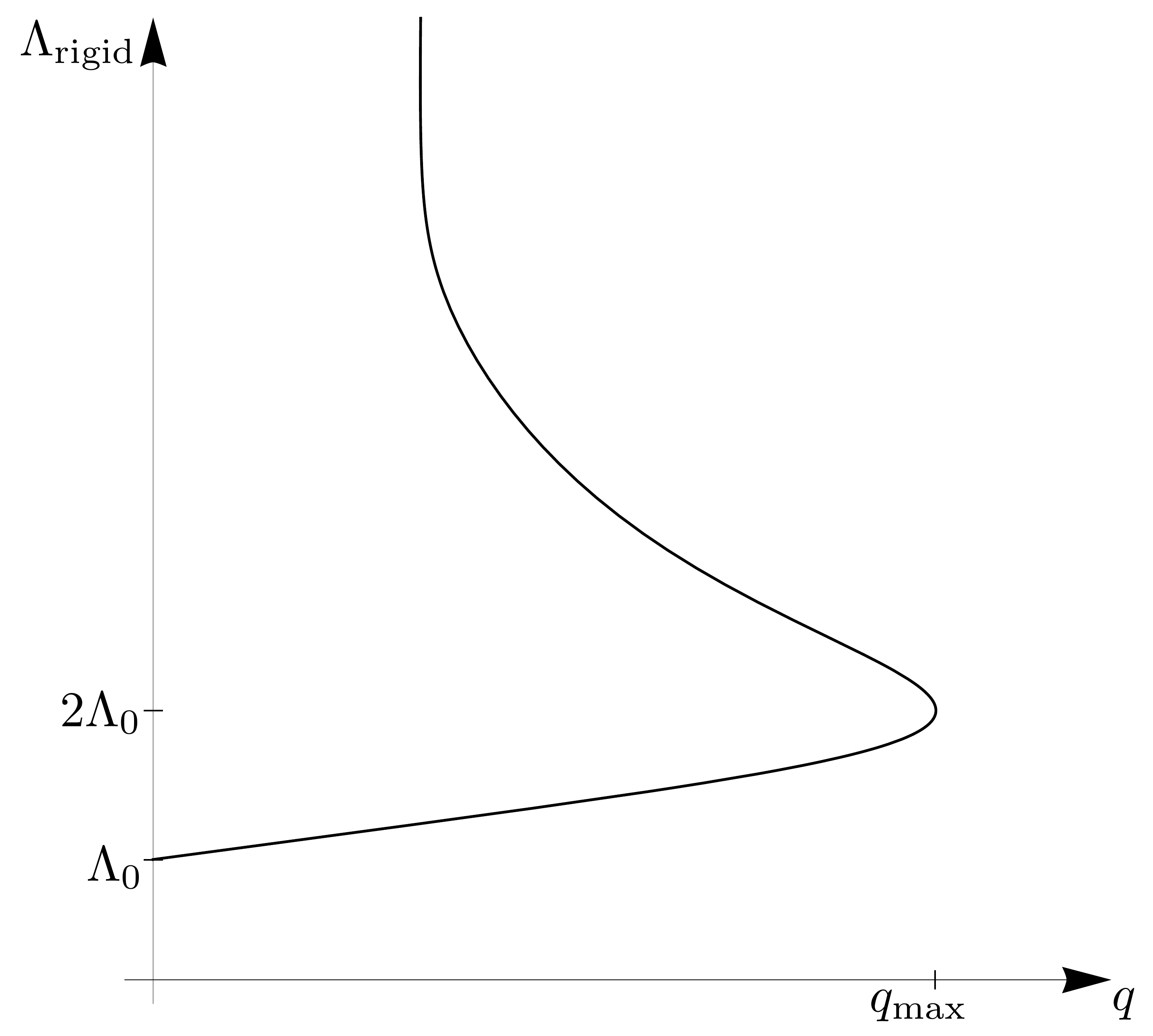}
\end{center}
\caption{The cosmological constant perceived by the $k=0$-physicits in dependence
on their natural RG scale $q$. While they are able to consistently
interpret the diagram's lower branch ($\Lambda_{0}\leq\Lambda_{{\rm rigid}}\leq2\Lambda_{0}$),
the ``scale horizon'' at $q_{{\rm max}}$ prevents them from passing
to the upper branch straightforwardly.\label{fig:figure-6-lambda-q-parametric}}
\end{figure}

Remarkably enough, this scale horizon has nothing to do with ``exotic''
Planck mass physics. Rather it occurs where the semiclassical $k^{4}$-running
of the cosmological constant becomes appreciable, namely at the much
lower turning point scale $k_{{\rm T}}$. Recall that if we were to
model real Nature by a Type IIIa trajectory, we find a turning
point scale as low as $k_{{\rm T}}\approx10^{-30}m_{{\rm Pl}}$, see
{[}h3{]}.\vspace{3mm}\\
{\bf (3)}
Figure \ref{fig:figure-6-lambda-q-parametric} illustrates the role
played by the horizon in connection with the cosmological constant
(problem). The reparametrized action $\Gamma_{q}$ of eq.~(\ref{eq:75-page78-EAA-via-g_0})
includes a cosmological constant term $\propto\Lambda_{k\left(q\right)}\int d^{4}x\sqrt{\bar{g}_{0}^{{\rm sc}}}$,
and the $k=0$-physicists regard $\Lambda_{k\left(q\right)}\equiv\Lambda_{{\rm rigid}}\left(q\right)$
as their natural scale dependent cosmological constant. Combining
eqs.~(\ref{eq:33-page62-def-Y}) and (\ref{eq:74-page77-q-square-via-k-square})
it is given by $\Lambda_{k\left(q\right)}=\Lambda_{0}Y_{k\left(q\right)}=\Lambda_{0}k^{2}\left(q\right)/q^{2}$,
and with (\ref{eq:107-page83-k_pm-branches-before-kT}) we obtain
the double-valued relation 
\begin{eqnarray}
\Lambda_{{\rm rigid}}\left(q\right) & = & 2\Lambda_{0}\left(\frac{q_{{\rm max}}}{q}\right)^{4}\left[1\pm\sqrt{1-\left(\frac{q}{q_{{\rm max}}}\right)^{4}}\right]\,.\label{eq:LR-page90-1-lambda-rigid-via-q}
\end{eqnarray}
This relation is depicted in Figure \ref{fig:figure-6-lambda-q-parametric},
with the plus (minus) sign corresponding to the upper (lower) branch
of the diagram. 

The $k=0$-physicists have no logical difficulties
interpreting the lower branch of $\Lambda_{{\rm rigid}}\left(q\right)$.
They are unable, however, to pass around the horizon at $q=q_{{\rm max}}$,
$\Lambda_{{\rm rigid}}\left(q\right)=2\Lambda_{0}$ if they insist
on using the scale $q$. Seen as a curve parameter which parametrizes
the RG trajectory, $q$ is a ``good'' coordinate on the RG time
axis only below the turning point. In order to go beyond the horizon
a ``better'' coordinate is needed, such as $k$ for example, which
is acceptable even globally. This hints at a certain analogy between
the scale horizon and the familiar coordinate horizons in spacetime.

On a more positive note we may conclude that nevertheless \emph{the
rigid picture based upon the perturbative, i.e., the $k_{-}$-branch
is applicable and equivalent to the running picture provided no relevant
momenta $q$ exceed $q_{{\rm max}}$.}

\subsection{The boundedness of $n_{{\rm COM}}\left(k\right)$ and $q\left(k\right)$}

While the scale $q$ is not a fully satisfactory alternative to $k$
as a RG time, the function $q=q\left(k\right)$ has another important,
and logically independent application, namely the determination of
the cutoff modes along the flow. This application does \emph{not}
require the inverse function.\vspace{3mm}\\{\bf (1)} In Subsection
\ref{sub:Cutoff-modes:-relating-nCOM-and-q} we showed that $n_{{\rm COM}}\left(k\right)$
is determined by eq.~(\ref{eq:76-page79-calF_0-via-q}), which requires
$q\left(k\right)$ as the essential input. Specializing for $S^{4}$-spacetimes,
eq.~(\ref{eq:77-page79-nCOM-via-q-of-k}) yields $n_{{\rm COM}}\left(k\right)=\bar{r}_{0}q\left(k\right)$,
which is valid in the approximation of a quasi-continuous spectrum
$\left(n\gg1\right)$ employed throughout. Thus, the $n$-quantum
number of the cutoff mode is known as an explicit function of $k\in\left[0,\infty\right)$:
\begin{eqnarray}
n_{{\rm COM}}\left(k\right) & = & \begin{cases}
\frac{\bar{r}_{0}k}{\left(1+\ell^{4}k^{4}\right)^{1/2}} & \mbox{if }0\leq k\apprle\hat{k}\\
\frac{\bar{r}_{0}}{L} & \mbox{if }\hat{k}\apprle k<\infty
\end{cases}\,.\label{eq:120-121-page91-nCOM-via-r0-and-k}
\end{eqnarray}
The scale dependent $n_{{\rm COM}}$ implies a corresponding decomposition
$\Upsilon_{{\rm UV}}\left(k\right)\cup\Upsilon_{{\rm IR}}\left(k\right)$
of all normal modes: modes with quantum numbers $n\geq n_{{\rm COM}}\left(k\right)$
belong to $\Upsilon_{{\rm UV}}\left(k\right)$, all others to $\Upsilon_{{\rm IR}}\left(k\right)$.

Since, for spheres, $n_{{\rm COM}}\left(k\right)$ differs from $q\left(k\right)$
by a constant factor only, Figure \ref{fig:figure-4-q-plot} can also
be regarded as a representation of $n_{{\rm COM}}$ in dependence
on the global RG parameter $k$. Therefore we conclude that $n_{{\rm COM}}\left(k\right)$
assumes a maximum at the turning point:
\begin{equation}
\boxed{
n_{{\rm COM}}\left(k\right)\,\leq\,n_{{\rm COM}}\left(k_{{\rm T}}\right) = \bar{r}_{0}\,q_{{\rm max}}\,=\,\bar{r}_{0}\,k_{{\rm T}}/\sqrt{2}\,,\,\mbox{ for all }k\in\left[0,\infty\right)\,.
} \label{eq:125-page92-nCOM-at-kT}
\end{equation}
The integer $n_{{\rm COM}}\left(k\right)$ is bounded above and never
becomes very large: \emph{nowhere along the Type IIIa trajectory,
evaluated ``on-shell'' with a self-consistent background, a cutoff
quantum number occurs} \emph{that would exceed the turning point value}.

It must be stressed that this result\footnote{Note that the impossibility of using $q$ as an alternative flow parameter
is irrelevant here.} is perfectly well-defined, conceptually meaningful, and in fact related
to a dynamical mechanism that is easily understood in general physical
terms, as we shall discuss below.\vspace{3mm}\\{\bf (2)} According
to Figure \ref{fig:figure-4-q-plot}, there can exist pairs of scales,
$k_{<}$ and $k_{>}$, smaller and larger than $k_{{\rm T}}$, respectively,
giving rise to the same quantum number $n_{{\rm COM}}\left(k_{<}\right)=n_{{\rm COM}}\left(k_{>}\right)$.
This proves, within the Einstein-Hilbert truncation, that the spectral
flow sketched schematically in Figure \ref{fig:figure-3-spectral-flow}
is indeed qualitatively correct. 

We mentioned already the possibility that certain eigenvalues ${\cal F}_{n}\left(k\right)$
intersect the diagonal \emph{twice}. At one scale they change their
UV/IR-status in the IR$\rightarrow$UV direction, while they move
in the opposite direction UV$\rightarrow$IR at another scale (indicated
in Figure \ref{fig:figure-3-spectral-flow} by the black and open
circles, respectively.)

Consistent with that, the plot in Figure \ref{fig:figure-4-q-plot}
reveals that $n_{{\rm COM}}\left(k\right)$ \emph{decreases}, implying
that the set $\Upsilon_{{\rm IR}}\left(k\right)$ looses modes, when
$k$ is increased further above $k_{{\rm T}}$. Actually this is the
same phenomenon which is also visible in Figure \ref{fig:figure-3-spectral-flow},
albeit in a different way: The cutoff mode for $k=k_{2}$ is identified
by that particular eigenvalue ${\cal F}_{n}\left(k\right)$ which
lies on, or just barely above the diagonal at $k=k_{2}$. Now, given
that ${\cal F}_{n}\left(k_{2}\right)\propto k_{2}^{4}$ grows very
rapidly with $k_{2}$, \emph{basically all eigenvalues }${\cal F}_{n}\left(k_{2}\right)$
\emph{will eventually exceed }$k_{2}^{2}$ when we let $k_{2}\rightarrow\infty$.
Hence $\Upsilon_{{\rm IR}}\left(k\right)$ looses more and more modes
when $k_{2}\rightarrow\infty$, and so $n_{{\rm COM}}\left(k_{2}\right)$
decreases correspondingly. 

The occurrence of this phenomenon is now fully confirmed by the explicit
Einstein-Hilbert result plotted in Figure \ref{fig:figure-4-q-plot},
where $n_{{\rm COM}}\left(k_{2}\rightarrow\infty\right)$ does indeed
decrease to a very small value. It is determined by the fixed point
properties ultimately.\footnote{The approximation of the quasi-continuous spectrum may become invalid
then.}

It should be clear now that this unfamiliar behavior is by no means
in conflict with the usual rule ``integrating out modes, i.e., making
$\Upsilon_{{\rm IR}}$ smaller, requires $k$ to be lowered.'' This
rule applies to the calculation of the $\Gamma_{k}\left[\varphi;\bar{g}\right]$
for $k$-independent off-shell arguments. Here instead we go on-shell
and follow the physical metrics $\bar{g}_{k}^{{\rm sc}}$ along the
RG trajectory.\vspace{3mm}\\{\bf (3)} The unexpected behavior
of $n_{{\rm COM}}\left(k\right)$, in particular its boundedness,
is one of our main results. For this reason let us emphasize that
the underlying mechanism is easily understood in elementary physical
terms and should be regarded particularly robust therefore.

The eigenvalues being of the form $\left(n/{\rm radius}\right)^{2}$,
we can make them larger in either of two ways, namely by increasing
$n$, or by decreasing the radius of the sphere. The first way is
the one we are familiar with from the off-shell EAA, leading to the
standard connection, 
$\left(\mbox{growing }k\right)\,\,\Leftrightarrow \left(\mbox{growing cutoff eigenvalue}\right)\Leftrightarrow\left(\mbox{growing }n\right)$. 

The second way, decreasing the radius, becomes an option only when
the background geometry is taken on-shell. But then it may happen
that a certain increment of $k$ is ``used up'' predominantly to
make the radius smaller, rather than to go to a higher quantum number.
What Figure \ref{fig:figure-4-q-plot} tells us is simply that, when
$k>k_{{\rm T}}$, the shrinking of the spacetime radius with growing
$k$ is so strong that we even can afford \emph{lowering} $n$ and
nevertheless get a \emph{bigger} ${\cal F}_{n}\left(k\right)$.

Thus it is also clear that the basic mechanism is not restricted to spheres
(having $r^{-2}\propto R\propto\Lambda$). All that is required is
a self-consistent geometry and a range of scales such that ${\cal F}_{n}\left(k\right)\propto\Lambda_{k}$.

\subsection{Spectral flow in a scaling regime}

\begin{figure}
\begin{center}
\includegraphics[scale=0.3]{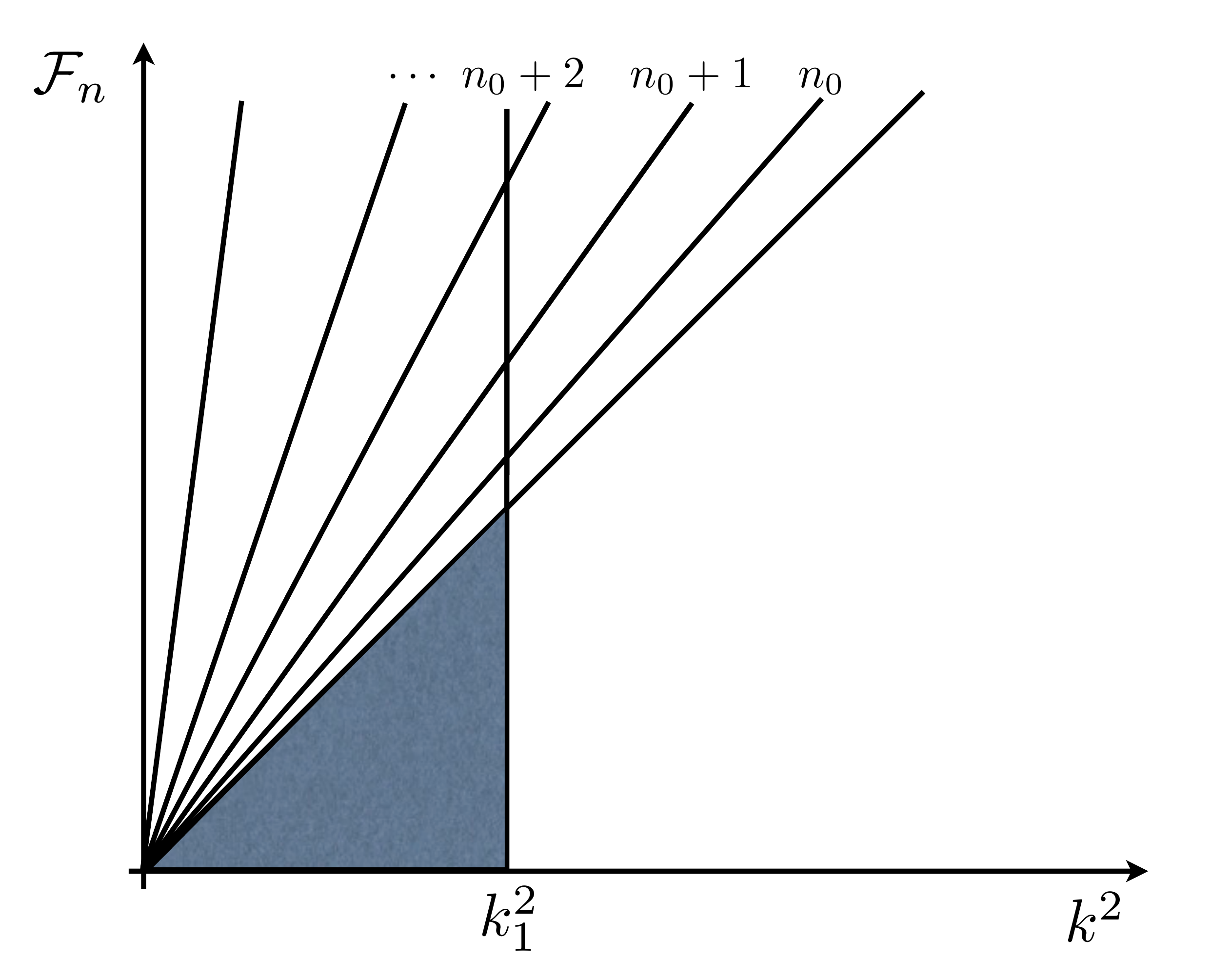}
\end{center}
\caption{Spectral flow of the ``undeformed'' fixed point theory.
Changing the value of $k_1$ does not lead to any exchange of modes between
$\Upsilon_{\rm{UV}}$ and $\Upsilon_{\rm{IR}}$.
\label{fig:figure-5}}
\end{figure}

In this paper we are mostly interested in the semiclassical regime
and in properties that are largely insensitive to the RG behavior
at $k\apprge\hat{k}=O\left(m_{{\rm Pl}}\right)$. Let us nevertheless
digress for a moment and assume that the RG trajectory is asymptotically
safe and hits a non-Gaussian fixed point when $k\rightarrow\infty$.
In its vicinity, $\Lambda_{k}=\lambda_{*}k^{2}$, and so all the eigenvalues
behave as ${\cal F}_{n}\left(k\right)\propto n^{2}k^{2}$, see Figure \ref{fig:figure-5}. 
Remarkably, no eigenvalues cross the ${\cal F}_{n}=k^{2}$-line
in this regime, and $\Upsilon_{{\rm IR}}\left(k\right)$ is the empty
set. 

Figure \ref{fig:figure-5} can be seen as the continuation for $k\rightarrow\infty$
of the flow in Figure \ref{fig:figure-3-spectral-flow} or, in its
own right, as the flow of the undeformed fixed point theory for all
$k\in\left[0,\infty\right)$.

\section{Application to the Cosmological Constant Problem \label{sec:Application-to-the-CC-problem}}

Finally let us critically reconsider the standard argument concerning
the alleged unnaturalness of a small $\Lambda$, which we reviewed
in the Introduction. As we shall see, this argument about the gravitational
effect of vacuum fluactuations is flawed by not giving due credit
to Background Independence. Instead, by doing so we can show that
the domain of validity of this calculation is considerably smaller
than expected, and that it has actually nothing to say about a potentially
large renormalization of $\Lambda$.

\subsection{Preliminaries and assumptions}

{\bf (1) The thought experiment.} Implicit in the reasoning of Subsection
\ref{sub:Summing-zero-point-energies} is the imaginary removal of
a selected quantum field, $Q\left(x\right)$ say, from the set of
all fields that exist in Nature, and the assumption that the remaining
fields jointly give rise to essentially the Universe as we know it.
One then gradually ``turns on'' the quantum effects of $Q\left(x\right)$
in this pre-exisiting Universe, and wonders about the backreaction
the extra field exerts on it.

The tacit assumption is that the backreaction is weak so that the
Universe has a chance to look like ours even with the extra field
fully quantized, but this then turns out not to be the case according
to the standard analysis. In our opinion this should first of all
raise a number of questions and concerns about the very setting of
this thought experiment, over and above the physics issues it tries
to address. For example, should the Universe before or after adding
$Q\left(x\right)$ be as large as ours? In the latter case, the Universe
without $Q\left(x\right)$ is strongly curved, so, why should the
calculation of $\Lambda$ in flat spacetime be sufficient? Under what
circumstances does the quantum Universe, with and without $Q\left(x\right)$,
actually appear to be semiclassical?

To fully avoid difficulties and conceptual problems of this kind we
replace the original thought experiment by a logically simpler and
more clear-cut question for theory. While aiming at the same phyics
issue, it avoids the dubious separation of a special field from the
rest of the Universe, and it also does not rely on the possibility
of treating one part of the Universe as classical, while the other
is quantum mechanical.

The question is as follows: \emph{In a fully quantum Universe with
all fields quantized, what is the shift of the cosmological constant,
$\Delta\Lambda$, that $k=0$-observers ascribe to the zero-point
energies of all quantum fields together?} Within the EAA approach
to quantum gravity, it will be possible to answer this question in
an unambiguous way, \emph{purely by inspection} of the trajectory,
$k\mapsto\Gamma_{k}$, and the trajectories derived from it, $k\mapsto\bar{g}_{k}^{{\rm sc}},\Upsilon_{{\rm IR}}\left(k\right)$.
\vspace{3mm}\\{\bf (2) Scope and assumptions.} Let us set up the
EAA framework now and outline its range of validity. The quantum fields
in questions are $h_{\mu\nu}$ and a collection of matter fieds $\psi\equiv\left(\psi_{\alpha}\right)$,
their dynamics being ruled by a given solution to the flow equation,
$\Gamma_{k}\left[h,\psi;\bar{g}\right]$. The argument we are going
to present is based upon the following assumptions then: \vspace{3mm}\\{\bf (i)}
We assume that, in the $G$-$\Lambda$-sector, the RG trajectory $k\mapsto\Gamma_{k}\left[h,\psi;\bar{g}\right]$
is qualitatively equivalent to a Type IIIa trajectory of the Einstein-Hilbert
truncation. Its $g$-$\lambda$ projection looks as in Figure \ref{fig:figure2-EH-flow-type-I-II-IIIa}b.

More precisely, it is sufficient that it does so for $k\apprle\hat{k}$.
While the existence of a turning point $\left(g_{{\rm T}},\lambda_{{\rm T}}\right)$
is of essential importance, our argument does not require a specific
$k\rightarrow\infty$ behavior, such as the approach of a fixed point,
for example. Referring back to Section \ref{sec:The-Einstein-Hilbert-example}
it is clear that the requirement of a turning point is met under very
general circumstances. All one needs is a semiclassical regime where
$\Lambda_{k}$ behaves as in eq.~(\ref{eq:T3-page49-1-dimful-lambda-linear-flow}),
i.e., $\Lambda_{k}\approx\Lambda_{0}+\nu G_{0}k^{4}$, whereby $\Lambda_{0}>0$
and $\nu>0$. \vspace{3mm}\\{\bf (ii)} Furthermore, we assume that
$G_{0}\Lambda_{0}\ll1$, i.e., $\Lambda_{0}\ll m_{{\rm Pl}}^{2}$,
as in real Nature. This condition is even weaker than in the standard
calculation: The latter sums up the zero-point energies of a field
\emph{in flat space} and so corresponds to letting $\Lambda_{0}=0$.

By (\ref{eq:T34-page49-10-ell-ellPl-L-ratios}) the condition $G_{0}\Lambda_{0}\ll1$
implies a clear separation of scales: $L\gg\ell\gg\ell_{{\rm Pl}}$.
Another consequence is that $n_{{\rm COM}}\left(k_{{\rm T}}\right)\gg1$,
which allows for the technically convenient approximation of a quasi-continuous
spectrum at the scales of interest. \vspace{3mm}\\{\bf (iii)} No
real matter particles are included. Virtual particles are taken into
account by the influence they have on the RG running of $G_{k}$ and
$\Lambda_{k}$. The $T_{k}^{\mu\nu}$-term in the tadpole equation
(\ref{eq:30-page60-Einstein-tadpole-eq}) is assumed negligible.\vspace{3mm}\\{\bf (iv)}
The solutions to the tadpole equation which we consider are $S^{4}$,
or de Sitter metrics of the rescaling-type (\ref{eq:32-page62-g-sc_k-via-Y-g_0}).

Because of the highly symmetric spacetime, the immediate applicability
of our discussion will be restricted to cosmology basically. Furthermore,
since the cosmological constant term in Einstein's equation must dominate
over $T_{k}^{\mu\nu}$, its natural domain of applications includes
\emph{a vacuum dominated era of late-time acceleration} such as
the one we presumably live in. Luckily, this is anyhow the regime
where the cosmological constant is observationally accessible to us.

The reader must be warned that, when interpreting our results,
it is important to keep the above limitations in mind and to refrain
from transferring the results to more complex physical problems involving
matter at a significant level, and/or less symmetric geometries with
several relevant scales. 

When the stress-energy tensor in Einstein's
equation is more important than the $\Lambda$-term, and is $k$-independent
in the regime of interest, the new on-shell effects resulting from
a running metric \emph{will disappear immediately}. 

Moreover, in multi-scale
problems a straightforward tree-level evaluation of the EAA is insufficient
in most cases. In particular, we do not expect that typical laboratory-scale 
scattering experiments are in any way affected by those effects
since the standard model interactions are overwhelmingly strong at
the corresponding energies. \vspace{3mm}\\{\bf (3) What will (not) be shown.}
It is to be stressed that we do \emph{not} claim, or try to prove,
that the solutions to the RG equations necessarily yield values of
$G_{0}\Lambda_{0}$ which are always small, thus explaining why Nature
could not but choose the exeedingly tiny number $G_{0}\Lambda_{0}\approx10^{-120}$.

Rather, we show that \emph{if} the Universe is described by a trajectory
having $G_{0}\Lambda_{0}\ll1$, \emph{then }$k=0$-physicists can
rightfully attribute an energy density to the quantum vacuum fluctuations
which is {\it at most of the same order as the $\Lambda_{0}$-contribution}.
Clearly, this result is quite different from the usual one, and importantly,
it cannot nurture any ideas about a small $\Lambda_{0}$ being ``unnatural''
in presence of quantum fields.

\subsection{No naturalness problem due to vacuum fluctuations}

Let us go through the various steps of the EAA-based reasoning now.\vspace{3mm}\\{\bf (1) The input.}
Starting out from the given RG trajectory, $k\mapsto\Gamma_{k}$,
we construct the associated trajectories in the spaces of metrics,
$k\mapsto\bar{g}_{k}^{{\rm sc}}$, and of Laplacians, $k\mapsto\Box_{\bar{g}_{k}^{{\rm sc}}}$.
The spectral flow of the latter then attaches well-defined sets $\Upsilon_{{\rm IR}}\left(k\right)$
to all points $\Gamma_{k}$ along the trajectory.\vspace{3mm}\\{\bf (2) One spectral flow only.}
In general the cutoff modes and $\Upsilon_{{\rm IR}}\left(k\right)$
would depend on the tensor type of the field $\Box_{\bar{g}_{k}^{{\rm sc}}}$
acts upon. In the case at hand we are entitled to ignore this dependence.
We are interested in the quasi-continuous part of the spectrum (quantum
numbers $n\gg1$) where the eigenvalues are independent of the tensor
rank, cf.~Subsection \ref{sub:S4-type-space}. (Their degeneracies
are not, but they play no role.)\footnote{Analogous remarks apply to fermions. For Dirac fields one may employ
the squared Dirac operator in place of the Laplacian. The two operators
differ by a non-minimal (curvature) term which, again, is inessential
for not too low eigenvalues.}\vspace{3mm}\\{\bf (3) Read off rather than invent: the degrees of freedom.}
At every fixed scale $k$, $\Upsilon_{{\rm IR}}\left(k\right)\equiv\left\{ \chi_{n}\big|n<n_{{\rm COM}}\left(k\right)\right\} $
is comprised of those modes that are not ``integrated out'' at the
point $\Gamma_{k}$ on the trajectory, i.e., their quantum fluctuations
are not yet accounted for by renormalized values of the coupling constants
in the EAA.

This allows us to conclude that $\Upsilon_{{\rm IR}}\left(k\right)$
can be regarded the precise description, and translation to the EAA
framework, of the degrees of freedom whose zero point energies are
summed up by the standard calculation in the Introduction. The modes
$\Upsilon_{{\rm IR}}\left(k\right)$ constitute a classical field
theory for which $k$, similar to ${\cal P}$ before, plays the role
of an \emph{ultraviolet} cutoff, and $\Gamma_{k}$ has a status analogous
to a bare action specified at this scale.\vspace{3mm}\\{\bf (4) The rigid picture comes into play.}
To complete the reinterpretation of the traditional $\rho_{{\rm vac}}$-computation
in the Background Independent setting we observe that this computation
must be ascribed to $k=0$\emph{-physicists} as it employs the
rigid and (almost) flat metric $\left(\bar{g}_{0}^{{\rm sc}}\right)_{\mu\nu}$
to a maximum extent. The dispersion relation $\omega\left({\bf p}\right)=\left|{\bf p}\right|$
and the cutoff $\left|{\bf p}\right|\leq{\cal P}$, for instance,
involve the flat metric. Hence, on the EAA-side, it is the ``rigid''
picture of the RG flow that must be used in a comparison of the standard
and the Background Independent approach. Therefore we re-express the
EAA at this stage in the form of the new action $\Gamma_{q}$ introduced
in Subsections \ref{sub:From-the-Running-to-rigid-picture} and \ref{sub:Completing-the-rigid-picture}.\vspace{3mm}\\{\bf (5) What the EAA tells us about $\Delta \Lambda$.}
Now we pose the question: Assume a team of $k=0$-physicists have
measured the cosmological constant of their Universe, $\Lambda_{0}$,
and they are in possession of the untruncated EAA functional.
They employ it in order to deduce the contribution to the cosmological
constant, $\Delta\Lambda$, that originates from the zero point oscillations
of all quantum fields. What order of magnitude will they find for
the ratio $\Delta\Lambda/\Lambda_{0}$?

The answer proceeds as follows: 
No matter what value of $k$ we pick, $\Upsilon_{{\rm IR}}\left(k\right)$
is always a comparatively ``small'' subset of all the field modes,
its elements having $-\Box_{\bar{g}_{k}^{{\rm sc}}}$-eigenvalues
that are bounded above and ``quantum numbers'' $n<n_{{\rm COM}}\left(k\right)$.
The function $n_{{\rm COM}}\left(k\right)$, sketched in Figure \ref{fig:figure-4-q-plot},
has a maximum at $k=k_{{\rm T}}$. Thus we conclude, on the basis
of the actual RG evolution, that the largest possible set of not-yet-quantized
degrees of freedom, $\Upsilon_{{\rm IR}}\left(k_{{\rm T}}\right)$,
occurs in the effective field theory belonging to the turning point, the hallmark of a Type IIIa trajectory.

Let us now fix some scale $k_{1}$ and read off the \emph{true} contribution
$\Delta\Lambda$ which, upon quantization, the modes in $\Upsilon_{{\rm IR}}\left(k_{1}\right)$
will supply to the cosmological constant. We perform the quantization
within the EAA framework where it amounts to RG-evolving the ``bare''
action $\Gamma_{k_{1}}$ down to the effective action $\Gamma\equiv\Gamma_{k=0}$
by following the prescribed trajectory. One of the running couplings
involved is the total cosmological constant, and it evolves from $\Lambda_{k_{1}}$
to $\Lambda_{0}$. As a result, $\Delta\Lambda=\Lambda_{k_{1}}-\Lambda_{0}$
is the shift due to the quantization of the modes in $\Upsilon_{{\rm IR}}\left(k_{1}\right)$.
For the simplified trajectory in eq.~(\ref{eq:T3-page49-1-dimful-lambda-linear-flow})
we obtain
\begin{eqnarray}
\Delta\Lambda & = & \nu G_{0}k_{1}^{4}\,.\label{eq:8-30-page156-Delta-Lambda}
\end{eqnarray}

At this point it becomes crucial to carefully distinguish the ``running''
and the ``rigid'' pictures of the RG evolution. \vspace{3mm}\\{\bf (a) Running picture.}
The running picture employs $k\in\left[0,\infty\right)$ as the independent
evolution parameter. Therefore it is meaningful to consider $\Upsilon_{{\rm IR}}\left(k\right)$
and the corresponding shift $\Delta\Lambda$ for any $k_{1}\in\left[0,\infty\right)$,
so that, by (\ref{eq:8-30-page156-Delta-Lambda}), $\Delta\Lambda\propto k_{1}^{4}$
can become arbitrarily large when $k_{1}$ is a scale sufficiently
high. The physics interpretation of this large $\Delta\Lambda$ will
be given in subsection \ref{sub:Where-has-all-large-curvature-gone}.\vspace{3mm}\\
{\bf (b) Rigid picture.}
Below the turning point, the description of the RG evolution in terms
of $\Gamma_{q}$, involving $\Lambda_{{\rm rigid}}\left(q\right)$,
with $q\in\left[0,q_{{\rm max}}\right]$, is equivalent to the running
picture. As we know, the function $q\left(k\right)\equiv n_{{\rm COM}}\left(k\right)/\bar{r}_{0}$
assumes a maximum at $k=k_{{\rm T}}$, and $q$ is bounded above,
$q\leq q_{{\rm max}}=q\left(k_{{\rm T}}\right)=k_{{\rm T}}/\sqrt{2}$,
cf.~Figure \ref{fig:figure-4-q-plot}. Inverting it locally, one
obtains two branches, $k=k_{\pm}\left(q\right)$, with $k_{\pm}\left(q_{{\rm max}}\right)=k_{{\rm T}}$,
cf.~(\ref{eq:107-page83-k_pm-branches-before-kT}), but we use only
the perturbative ``minus''-branch on which $q=0\Leftrightarrow k=0$.

In order to determine $\Delta\Lambda$, the ``$k=0$-physicists''
adhering to the rigid picture must first of all fix a certain momentum,
$q_{1}$, in such a way that this $q_{1}$ is actually realized as
a parameter value that specifies a point $\Gamma_{q_{1}}$ on the
RG trajectory given. This implies that a scale $q_{1}\in\left[0,q_{{\rm max}}\right]$
must be chosen, since no value $k\in\left[0,\infty\right)$ of the
global RG parameter will ever result in a scale $q>q_{{\rm max}}$.

Given $q_{1}\leq q_{{\rm max}}$, there exists a well defined
set of not-yet-quantized modes $\Upsilon_{{\rm IR}}\left(k_{-}\left(q_{1}\right)\right)$,
and we ask about the shift $\Delta\Lambda$ caused by the inclusion
of their fluctuations. On the perturbative branch, $q_{1}\rightarrow0$
amounts to $k_{-}\left(q_{1}\right)\rightarrow0$, and so $\Delta\Lambda=\Lambda_{k_{-}\left(q_{1}\right)}-\Lambda_{k_{-}\left(0\right)}=\Lambda_{{\rm rigid}}\left(q_{1}\right)-\Lambda_{0}$
with $\Lambda_{{\rm rigid}}$ defined as in Subsection \ref{sub:Scale-horizon-and-failure-of-q-picture}.
The explicit function $\Lambda_{{\rm rigid}}\left(q\right)$ was given
in eq.~(\ref{eq:LR-page90-1-lambda-rigid-via-q}), and Figure \ref{fig:figure-6-lambda-q-parametric}
shows a plot of it.

Along the perturbative branch, the lower one in Figure \ref{fig:figure-6-lambda-q-parametric},
$\Lambda_{{\rm rigid}}\left(q_{1}\right)$ grows monotonically by
a factor of $2$ when the $k=0$-physicist increase the scale from
$q_{1}=0$ to $q_{1}=q_{{\rm max}}$, changing from $\Lambda_{{\rm rigid}}\left(0\right)=\Lambda_{0}$
to $\Lambda_{{\rm rigid}}\left(q_{{\rm max}}\right)=2\Lambda_{0}$.
The shift $\Delta\Lambda$ is maximum when $q_{1}=q_{{\rm max}}$,
yielding $\Delta\Lambda=\Lambda_{{\rm rigid}}\left(q_{{\rm max}}\right)-\Lambda_{0}=\Lambda_{0}$,
and so
\begin{equation}
\boxed{
\frac{\Delta\Lambda}{\Lambda_{0}} = 1\,.
} \label{eq:8-40-page161-DeltaLamdbda-over-Lambda0}
\end{equation}

This, finally, is the answer to the question we posed above: \emph{Within
the rigid picture, when applicable, the shift of the cosmological
constant due to the vacuum fluctuations will always be
at most of the order of magnitude of the IR value $\Lambda_{0}$. }

This completes the proof of the assertion made above. \vspace{3mm}\\{\bf (6) To be, or not to be applicable.}
The above answer subsumes two logical possibilities:\vspace{3mm}\\{\bf (i)}
The rigid picture is applicable and implies $\Delta\Lambda\leq\Lambda_{0}$.
This requires all $q$, in particular the UV cutoff $q_{1}$ for $\Upsilon_{{\rm IR}}\left(k_{-}\left(q_{1}\right)\right)$,
to be not greater than $q_{{\rm max}}$. So, since $q_{1}$ is equivalent
to the cutoff ${\cal P}$ from the Introduction, ${\cal P}\equiv q_{1}\leq q_{{\rm max}}$.\vspace{3mm}\\{\bf (ii)}
The rigid picture is not applicable as we compute the $\Delta\Lambda$
for modes $\Upsilon_{{\rm IR}}\left(k_{1}\right)$ with $k_{1}$ beyond
the scale horizon: $k_{1}>k_{{\rm T}}=k_{-}\left(q_{{\rm max}}\right)$.
In this case, $\Delta\Lambda\propto k_{1}^{4}$ can be large, but
the effective theories for $k=0$ and $k=k_{1}$, respectively, are
no longer continuously connected by a RG trajectory $q\mapsto\Gamma_{q}$
whereby $q$ is the proper-momentum of a scale independent metric
$\bar{g}_{0}^{{\rm sc}}$. They lie on opposite sides of a ``scale
horizon''.

In the following subsections we discuss the two cases in turn.

\subsection{Validity of the standard calculation \label{sub:To-what-extent-sum-0-energies-meaningful}}

Now we are able to pinpoint what precisely is wrong about the standard
arguments concluding that huge values of summed-up zero point energies
render a small $\Lambda_{0}$ unnatural. In the present notation the
integral (\ref{eq:1-1-page6-rho_vac-flat-space}) reads
\begin{eqnarray}
\rho_{{\rm vac}} & \equiv & \frac{\Delta\Lambda}{8\pi G_{0}}\,=\,\frac{1}{2}\hbar\int_{\left|{\bf q}\right|\leq{\cal P}=q_{1}}\frac{d^{3}q}{\left(2\pi\right)^{3}}\left|{\bf q}\right|\,=\,cq_{1}^{4}\,.\label{eq:8-50-page163-rho-vac_q1-cutoff-flat-space}
\end{eqnarray}
We switched to the notation ${\bf q}$ for the momentum since it is
``proper'' with respect to a scale independent (almost flat) metric,
$\left(\bar{g}_{0}^{{\rm sc}}\right)_{\mu\nu}\approx\eta_{\mu\nu}$,
and the same metric underlies the kinetic term of the massless free
field that is being quantized.

Let us now estimate the domain of applicability of (\ref{eq:8-50-page163-rho-vac_q1-cutoff-flat-space})
by trying to identify it with a sub-sector of a Background Independent
EAA analysis based on a trajectory $k\mapsto\left\{ \Gamma_{k},\bar{g}_{k}^{{\rm sc}}\right\} $
describing an approximately free massless field at low scales, the
graviton being the prime example.

At least in principle, the classical $\Upsilon_{{\rm IR}}\left(k_{1}\right)$-theory,
once identified by means of the EAA, can also be quantized by any
other, that is, non-FRG method. In this spirit, (\ref{eq:8-50-page163-rho-vac_q1-cutoff-flat-space})
can be seen as an alternative, albeit approximate way of recovering
the cosmological constant's leading RG running, \emph{provided one
identifies the UV cutoff ${\cal P}$ with the floating RG scale}.
Concretely, ascribing (\ref{eq:8-50-page163-rho-vac_q1-cutoff-flat-space})
to the rigid picture, we must identify ${\cal P}=q_{1}$, as indicated
in (\ref{eq:8-50-page163-rho-vac_q1-cutoff-flat-space}).

However, when it comes to interpreting the role of the ${\cal P}=q_{1}$-dependence
in (\ref{eq:8-50-page163-rho-vac_q1-cutoff-flat-space}), the Background
Independent approach (EAA) and the standard one (QFT in classical
curved spacetime) differ in a significant way.\vspace{3mm}\\{\bf (8a) Standard approach:}
In traditional QFT on a fixed spacetime, one starts by making the
divergent integral finite in an arbitrary way which needs not to have
a physical interpretation, here by a momentum cutoff, then one adds
to $\Delta\Lambda\equiv\Delta\Lambda\left({\cal P}\right)$ a ${\cal P}$-dependent
counterterm $\Lambda_{{\rm ct}}\left({\cal P}\right)$ such that $\Delta\Lambda\left({\cal P}\right)+\Lambda_{{\rm ct}}\left({\cal P}\right)$
approaches the observed cosmological constant $\Lambda_{{\rm obs}}$
for ${\cal P}\rightarrow\infty$, and only when this limit is taken,
once and only once, one computes a metric, solving Einstein's equation
with the parameter $\Lambda_{{\rm obs}}$ inserted.\footnote{Clearly, this rough description can be refined or modified in many
ways; they would not affect the main conclusion though.} Hereby, the adjustment of $\Lambda_{{\rm ct}}\left({\cal P}\right)$
involves the notorious fine-tuning which is behind the naturalness
aspect of the cosmological constant problem.

Note that, in the standard setting, the only point of contact between
the QFT calculation and the curvature of spacetime is the single quantity
$\Lambda_{{\rm obs}}$. In the language of perturbative renormalization
theory, it is one of the ``renormalized'' or ``physical'' parameters
which appear in the conventional effective action $\Gamma$, with
the UV cutoff removed, and in absence of any IR cutoff. While this
allows us to identify $\Lambda_{{\rm obs}}$ with $\Lambda_{k=0}$
from the EAA side, $\Lambda_{{\rm obs}}\equiv\Lambda_{0}$, it also
points towards the following potential insufficiency of the standard
treatment which we have not mentioned yet: 

Whatever the vacuum fluctuation is like, no matter how violent it
is, or whether it has a Planck- or a Hubble- size wavelength, it
can impact the geometry of spacetime only via the eye of a needle,
namely the renormalized cosmological constant, $\Lambda_{0}$. In
Einstein's equation it multiplies the no-derivative term and controls
the properties of solutions at the largest possible length scales.
But is it really plausible to assume that, say, micrometer-scale fluctuations
influence the spacetime geometry predominantly on \emph{cosmological}
scales? Why should cause and effect be separated by 30 orders of magnitude?
We shall come back to this particular aspect of the problem in Subsection
\ref{sub:Where-has-all-large-curvature-gone}.\vspace{3mm}\\{\bf (8b) EAA approach:}
There are two relevant characteristics. First, the Background Independence
of the setting allows the dynamics to determine the background metric,
and second, every quantum field theory is regarded as the limit $k\rightarrow0$
of a sequence of effective field theories. The combination of both
properties implies that, unavoidably, dynamically selected backgrounds
are scale dependent. According to the effective field theory interpretation
of $\Gamma_{k}$ and $\Gamma_{q}$, the integral (\ref{eq:8-50-page163-rho-vac_q1-cutoff-flat-space})
describes how the cosmological constant, and via (\ref{eq:30-page60-Einstein-tadpole-eq})
the curvature of spacetime, depend on the momentum $q$ of the probe
that is used to explore the structure of the Universe. The full quantum
theory is obtained by letting $k\approx q\rightarrow0$.

By comparing {\bf (8a)} and {\bf (8b)} it becomes evident that,
for no Type IIIa trajectory $k\mapsto\left\{ \Gamma_{k},\bar{g}_{k}^{{\rm sc}}\right\} $
whatsoever, the standard background dependent calculation can be equivalent
to, or a meaningful approximation of the Background Independent EAA-based
one.

In the standard approach, the $q_{1}={\cal P}\rightarrow\infty$ limit
of the integral (\ref{eq:8-50-page163-rho-vac_q1-cutoff-flat-space})
should be taken, while the EAA requires $q_{1}\rightarrow0$. The
problem here is not that the limits are different, and that the limit
does not exist in the former case: After all we are prepared to admit
a counterterm. What is fatal for the standard approach is rather that
the entire flat space-based calculation leading to (\ref{eq:8-50-page163-rho-vac_q1-cutoff-flat-space})
breaks down already at a scale much lower than ${\cal P}$, namely
$q_{1}=q_{{\rm max}}=k_{{\rm T}}/\sqrt{2}$. From there on the rigid
picture cannot be mantained any longer. It is then advisable to switch
to the running picture by means of the which the horizon can be crossed
without problems. 

As long as the calculation is valid, the summed zero point energies
given by (\ref{eq:8-50-page163-rho-vac_q1-cutoff-flat-space}) never
exceed $\rho_{{\rm vac}}=cq_{{\rm max}}^{4}$, and $\Delta\Lambda$
is at most $\Delta\Lambda=8\pi cG_{0}q_{{\rm max}}^{4}=2\pi cG_{0}k_{{\rm T}}^{4}=\left(2\pi c/\nu\right)k_{{\rm T}}^{4}$.
So we see that
\begin{eqnarray}
\frac{\Delta\Lambda}{\Lambda_{0}} & = & O\left(1\right)\,,\label{eq:8-61-page169-DeltaLambda-over-Lambda}
\end{eqnarray}
which is essentially the same as (\ref{eq:8-40-page161-DeltaLamdbda-over-Lambda0}).

Thus, again, we conclude that \emph{the summation of zero point energies
in flat space cannot be used in order to claim that a small cosmological constant
is ``unnatural''}.

\subsection{Where has the large spacetime curvature gone? \label{sub:Where-has-all-large-curvature-gone}}

Above, in {\bf (5)} and {\bf (6)} of Subsection \ref{sub:To-what-extent-sum-0-energies-meaningful},
we mentioned that when one crosses the scale horizon and explores
scales $k_{1}>k_{{\rm T}}$, now relying on the running picture, the
induced cosmological constant can become arbitrarily large. In the
semiclassical regime, for example, $\Delta\Lambda\propto k_{1}^{4}$.

Now, this appears to be a resurrection of the notorious standard result
$\rho_{{\rm vac}}\propto{\cal P}^{4}$ with its disastrous consequence
that the natural size (Hubble length) of the Universe should be of
the order of a Planck length. This interpretation is false, however: 

By going through the steps of the EAA-based analysis we have learned
that, \emph{for a physical interpretation of the unbounded growth
of $\Delta\Lambda\left(k\right)$ above the turning point, it is compulsory
to equip spacetime with a running self-consistent metric, $\bar{g}_{k}^{{\rm sc}}$.
}This metric entails a scale dependent curvature then,
\begin{eqnarray}
R\left(\bar{g}_{k}^{{\rm sc}}\right) & = & 4\Lambda_{k}\,\approx\,4\Delta\Lambda\left(k\right)\,\propto\,k^{4}\,,\label{eq:8-100-page171-R-via-Lambda_k}
\end{eqnarray}
and according to the effective field theory interpretation, this curvature
is observed in experiments which scan spacetime with a probe involving
typical momenta of order $k$.

We assume that the RG trajectory has been selected such that $R\left(\bar{g}_{0}^{{\rm sc}}\right)=4\Lambda_{0}$
equals the curvature observed on the largest possible distance scale,
i.e.~at $k=0$. If we then increase $k$, \emph{the growing $\Lambda_{k}$
gives rise to growing curvature on smaller length scales of order
$k^{-1}$, while leaving unchanged the curvature measured on the largest,
i.e., cosmological scale}.

Even though at the technical level the behavior $\Delta\Lambda\propto{\cal P}^{4}$
of the integral (\ref{eq:8-50-page163-rho-vac_q1-cutoff-flat-space})
has the same origin as the $\Lambda_{k}\propto k^{4}$ running displayed
by the $\Gamma_{k}$ trajectory, the interpretations differ again
quite substantially: 

The effective field theory distributes the curvature
over many different scales and ascribes the ever growing curvature
$R=4\Lambda_{k}\propto k^{4}$ to the image of spacetime which is seen under
a ``microscope of resolving power $1/k$''. The standard procedure
summarized in {\bf (8a)}, Subsection \ref{sub:To-what-extent-sum-0-energies-meaningful},
on the other hand, cannot but associate the entire induced vacuum energy density
with the curvature on the largest cosmological scales. 

This is precisely
the insufficiency of the standard treatment we mentioned towards the
end of {\bf (8a)}, Subsection \ref{sub:To-what-extent-sum-0-energies-meaningful}.

It will be an interesting challenge for the future research to confirm
this ``multi-fractal'' character of spacetime \cite{Lauscher:2005qz,Reuter:2011ah,Rechenberger:2012pm,Calcagni:2013vsa}
by explicit calculations that do not rely on the effective field theory
interpretation \cite{pagani-zanusso}.\footnote{
See also ref.~\cite{Carlip:2018zsk} for an essentially classical discussion of small scale curvature hiding the cosmological constant.}
It would also be interesting to study the extension of the present work to 
include the RG flow of non-metric theories of various types
\cite{Daum:2010qt,Daum:2013fu,Pagani:2013fca,Pagani:2015ema,Reuter:2015rta,Manrique:2011jc,
Biemans:2016rvp,Biemans:2017zca,Houthoff:2017oam}.

\section{Summary \label{sec:Summary-and-conclusion}}

By its very definition, the gravitational average action $\Gamma_{k}\left[h_{\mu\nu},\psi,\cdots;\bar{g}_{\mu\nu}\right]$
is a functional that depends on $k$-independent fields over spacetime.
However, typical applications involve evaluating it, and its derivatives
at solutions of the effective field equations, and these (``on-shell'')
field configurations do depend on the RG scale.\vspace{3mm}\\
{\bf (1)}
In this paper we investigated consequences of this second type of $k$-dependence
which arises over and above the functional's explicit scale dependence.
While the latter follows directly from the functional RG equation
and has been studied in considerable detail during the past two decades,
the extra scale dependence stemming from field configurations
taken on-shell, and background metrics adjusted self-consistently,
is a largely unexplored territory yet.

It is clear though that also this second $k$-dependence must be taken
into account with care when it comes to matching the predictions of
$\Gamma_{k}$ against the real world. For the purposes of particle
physics on a non-dynamical spacetime one routinely computes average
actions like $\Gamma_{k}\left[\psi\right]$ which encode masses, coupling
constants, and similar properties of matter fields when gravity plays
no role. If the scope of the description is then extended to include
gravity, on the particle physics side the corresponding $\Gamma_{k}\left[h,\psi,\cdots;\bar{g}_{\mu\nu}\right]$
predicts again $k$-dependent masses and couplings. However, those
properties then pertain to elementary particles propagating \emph{on
a quite specific geometry}, namely a geometry whose metric is self-consistent
at precisely the scale of the running action, $k$. 

As we saw repeatedly in this paper, failure to correctly identify
this metric can lead to considerable errors and misconceptions, in
particular in view of the truly enormous scale differences that lurk
behind the very fast RG running of the cosmological constant.\vspace{3mm}\\{\bf (2)}
Both for technical simplicity and in order to amplify the unusual
new effects, throughout this paper we considered gravity \emph{in
absence of real matter particles}. Virtual, i.e., vacuum effects due
to in principle arbitrary matter fields were included though. Furthermore,
spacetime was assumed to be maximally symmetric in the explicit calculations,
which then restricts the immediate applications of the results to
the vacuum dominated era of cosmology. The conceptual developments
are valid much more generally. \vspace{3mm}\\{\bf (3)} We introduced
and applied a number of tools for extracting physics information from
the EAA which arises only after going on-shell, or as in our case,
by choosing the background self-consistently and setting the fluctuations
to zero. The discussion focussed on the eigenvalue equation of the
background Laplacian which, when still off-shell, organizes the coarse
graining and ``integrating out'' of field modes that underlies the
functional RG. We saw that upon letting $\bar{g}_{\mu\nu}\rightarrow\left(\bar{g}_{k}^{{\rm sc}}\right)_{\mu\nu}$
the eigenvalue equation turns into a complicated nonlinear relationship
between the quantum number characteristic of a mode's ``fineness''
and the RG parameter $k$. This relation is particularly striking
and counter-intuitive for trajectories of Type IIIa, the reason being
their turning point of the dimensionless cosmological constant. Increasing
$k$ above the turning point no longer leads to a finer, more structured
cutoff mode function, but rather brings one back to coarser ones with
a lower ``principal quantum number''. We explained and interpreted
this phenomenon in terms of the spectral flow along the generalized
RG trajectory $k\mapsto\left\{ \Gamma_{k},\bar{g}_{k}^{{\rm sc}}\right\} $,
which also provided us with a precise description of the not-yet-quantized
degrees of freedom, viz. the spaces $\Upsilon_{{\rm IR}}\left(k\right)$.\vspace{3mm}\\
{\bf (4)}
We observed and explained
the phenomenon of $\Upsilon_{{\rm IR}}\left(k\right)$ \emph{loosing}
modes when $k$ is increased. Actually it did not come quite unexpected. In \cite{Reuter:2005bb}
it has been shown that the effective spacetimes implied by the EAA
are similar to a ``fuzzy sphere'' whose degree of fuzzyness depends on $\lambda_k$. 
Even though the reasoning in \cite{Reuter:2005bb} is quite different from
the present one, it can be shown that they describe two faces of the
same medal. According to \cite{Reuter:2005bb}, the fuzzyness of spacetime, i.e.,
the impossibility to distinguish points that are too close, can also
be characterized by a minimum possible length, which has a subtle
interpretation though, see \cite{Reuter:2005bb,Reuter:2006qh,Reuter:2006zq}.\vspace{3mm}\\
{\bf (5)}
We exploited information about $\Upsilon_{{\rm IR}}\left(k\right)$
in order to contrast the conventionally used ``running picture''
of the trajectory $k\mapsto\left\{ \Gamma_{k},\bar{g}_{k}^{{\rm sc}}\right\} $
with a new one, the ``rigid picture'', which is more relevant from
the practical point of view. It corresponds to the situation human
particle physicists are in who perform laboratory-scale experiments
and, in their theoretical analysis, ``transform away'' (quantum)
gravity as far as possible. \vspace{3mm}\\{\bf (6)} As an application
of the rigid picture, we critically reassessed the status enjoyed
by the integrated zero-point energy of quantum fields, which frequently
is claimed to present a colossal threat to a value of the cosmological
constant as small as the one observed in real Nature, being roughly
$\Lambda\approx10^{-120}m_{{\rm Pl}}^{2}$. The EAA based, hence Background
Independent, treatment in Section \ref{sec:Application-to-the-CC-problem}
revealed that the traditional argument, while formulated within the
rigid picture, is flawed by the fact that the rigid picture in the
form used exists only over a rather short span of scales: A fluctuation-induced
change of $\Lambda$ by about a factor of $2$ is described consistenly,
but it is quite impossible to bridge 120 orders of magnitude by a
background dependent calculation on flat space. The rigid picture
breaks down already at the trajectory's turning point, i.e., in Nature
at the milli-electronvolt scale if matter does not intervene.

We concluded that \emph{it is not legitimate to interpret the standard,
and typically huge integrated zero point energies by claiming that
a small value of the cosmological constant is afflicted by a naturalness-
or finetuning-problem}.\vspace{3mm}\\{\bf (7)} Above the turning
point scale, use of the running picture is mandatory. There the Background
Independence built into the EAA framework allows the spacetime metric
to re-adjust continually during the RG evolution $k\mapsto\left\{ \Gamma_{k},\bar{g}_{k}^{{\rm sc}}\right\} $.
This leads to adiabatic changes of all excitation energies (eigenvalues)
in response to shrinking or expanding geometries, a possibility that
is unavailable in the traditional treatment. When $k$ is increased
beyond the turning point scale, $\Lambda_{k}$ and the corresponding
curvature may become large according to the running picture. In this
regime another deficiency of the standard treatment becomes relevant:
It is too simplistic in that all vacuum fluctuations cannot but contribute
to the curvature of the Universe \emph{on cosmological scales}. The
Background Independent RG approach suggests instead that fluctuations
of a given scale curve spacetime \emph{on that particular scale}. 

Thus, since we are able to measure $\Lambda$ on cosmological scales
only, at least for the time being, the overwhelming part of the vacuum
fluctuations might not have had a chance yet to manifest themselves
gravitationally in our observations and experiments. They could curve
spacetime on sub-cosmological length scales. If so, this resolves
the perhaps most mysterious aspect of the cosmological constant problem,
the question about the absence, or better, \emph{invisibility} of
substantial spacetime curvature attributable to vacuum fluctuations.
In any case, within the present analysis we do not find any tension,
let alone a ``clash'' between our theoretical expectations and actual
observations.

\end{spacing}


\end{document}